\documentclass[twocolumn,pre,superscriptaddress,bibnotes,notitlepage,nofootinbib,eqsecnum]{revtex4-2}

\usepackage{CJK}

\usepackage{graphicx}

\usepackage{dcolumn}

\usepackage{amsmath}

\numberwithin{equation}{section}

\usepackage{bm}
\usepackage[normalem]{ulem}
\usepackage{color}
\usepackage{amsmath}
\usepackage{amssymb}
\usepackage{lipsum}
\usepackage{bm}
\usepackage{xcolor}
\usepackage{graphicx}
\usepackage{verbatim}
\usepackage[T1]{fontenc}
\usepackage{hyperref}
 \def\rf#1{(\ref{#1})}
\newcommand{\bof}{\boldsymbol{\xi}_p}
\newcommand{\bor}{\boldsymbol{\xi}_\rho}

\usepackage{epstopdf}

\newcommand{\dep}{\delta p}
\newcommand{\dr}{\delta\rho}

\newcommand{\bp}{\mathbf{p}}
\newcommand{\bq}{\mathbf{q}}
\newcommand{\bv}{\mathbf{v}}

\newcommand{\br}{\mathbf{r}}

\newcommand{\bu}{\mathbf{u}}

\newcommand{\hn}{\hat{{\bf n}}}

\newcommand{\sep}{ \ \ \ , \ \ \ }

\newcommand{\beq}{\begin{equation}}
\newcommand{\eeq}{\end{equation}}
\newcommand{\beqn}{\begin{eqnarray}}
\newcommand{\eeqn}{\end{eqnarray}}
\newcommand{\pp}{\partial}

\newcommand{\la}{\langle}

\newcommand{\ra}{\rangle}

\newcommand{\vnab}{{\bf \nabla}}

\newcommand{\bew}{\begin{widetext}}
\newcommand{\ew}{\end{widetext}}
\newcommand{\nn}{\nonumber}

\begin{document}
\title{
{The inconvenient truth about flocks}}
\author{Leiming Chen}
\email{leiming@cumt.edu.cn}
\affiliation{School of Material Science and Physics, China University of Mining and Technology, Xuzhou Jiangsu, 221116, P. R. China}
\author{Patrick Jentsch}
\email{patrick.jentsch@embl.de}
\affiliation{Cell Biology and Biophysics Unit, European Molecular Biology Laboratory, 
Meyerhofstra\ss e1, 69117 Heidelberg, Germany}
\author{Chiu Fan Lee}
\email{c.lee@imperial.ac.uk}
\affiliation{Department of Bioengineering, Imperial College London, South Kensington Campus, London SW7 2AZ, U.K.}
\author{Ananyo Maitra}
\email{nyomaitra07@gmail.com}
\affiliation{{Laboratoire de Physique Th\'eorique et Mod\'elisation, CNRS UMR 8089, CY Cergy Paris Universit\'e, F-95032 Cergy-Pontoise Cedex, France}}
\affiliation{Laboratoire Jean Perrin, Sorbonne Universit\'{e} and CNRS, F-75005, Paris, France}
\author{Sriram Ramaswamy}
\email{sriram@iisc.ac.in}
\affiliation{Centre for Condensed Matter Theory, Department of Physics, Indian Institute of Science, Bangalore 560 012, India}
\affiliation{International Centre for Theoretical Sciences, Tata Institute of Fundamental Research, Bangalore 560 089 India}
\author{John Toner}
\email{jjt@uoregon.edu}
\affiliation{Department of Physics and Institute of Theoretical
Science, University of Oregon, Eugene, OR $97403^1$}
\affiliation{Max Planck Institute for the Physics of Complex Systems, N\"othnitzer Stra\ss e 38, 01187 Dresden, Germany}


\begin{abstract}
{We reanalyze the hydrodynamic theory of ``flocks''  that is, polar ordered ``dry'' active fluids in two dimensions. For ``Malthusian'' flocks, in which birth and death cause the density to relax quickly, thereby eliminating density as a hydrodynamic variable, we are able to obtain two exact scaling laws relating the three scaling exponents characterizing the long-distance properties of these systems. We also {show that it is highly plausible} that such flocks display long-range order in two dimensions.  In addition, we demonstrate that for ``immortal'' flocks, in which the number of flockers is conserved, the extra non-linearities allowed by the presence of an extra slow variable (number density) make it impossible to obtain any exact scaling relations between the exponents. We thereby demonstrate that several past published claims of exact exponents for Malthusian and immortal flocks are all incorrect.}
\end{abstract}

\maketitle

\section{Introduction}{\label{Intro}}

Flocks \cite{vicsek,TT95,TT98,John_bk} are a spectacular and ubiquitous example of spontaneously broken rotation invariance in active matter \cite{RMP, Annals, SRRev, chate2020dry}.  Specifically,  polar ordered dry (i.e., non-momentum conserving)  active fluids, unlike their equilibrium counterparts---planar magnets \cite{Mermin-Wagner,Hohenberg}--- can exhibit long-range orientational order in two dimensions. \cite{Toner_rean,chate2020dry}.
However,
an exact analytical theory of their large-distance, long-time scaling properties \cite{TT95,Annals,John_bk} has proved elusive \cite{Toner_rean}.
Such an exact calculation {in two dimensions} was believed possible for 
the  ``Malthusian'' \cite{Toner_mal} case, in which birth and death 
{suppress} density fluctuations, {making the density}{,} a fast, non-hydrodynamic variable. 

{In this article, we will reanalyze the hydrodynamic theory of both Malthusian and  ``immortal'' (that is, number-conserving) two-dimensional flocks.  We focus our attention on the three scaling exponents that 
characterize the long-distance,  long-time behaviour of these systems. These exponents are:
the ``dynamical'' exponent $z$ giving the scaling of time $t$ with distance $y$ in the direction perpendicular to the direction of mean flock motion, via the relation
\beq
t\propto |y|^z \,,
\label{zdef0}
\eeq
the anisotropy exponent $\zeta$ giving the scaling of distances $x$ along the direction of mean flock motion with distances along the 
direction perpendicular to the direction of mean flock motion, via the relation 
\beq
x\propto |y|^{
\zeta} \,,
\label{zetadef0}
\eeq
and the ``roughness'' exponent $\chi$ giving the scaling of the typical directional fluctuations $\delta\theta$ with distance $y$ in the 
direction perpendicular to the direction of mean flock motion, via the relation
\beq
\delta\theta\propto |y|^{
\chi} \,.
\label{chidef0}
\eeq

For Malthusian flocks, we derive two exact scaling relations between these three exponents:
\beq
\chi-\zeta+1=0 \,,
\label{1stscaleint}
\eeq
\beq
z-\zeta-2\chi-1=0 \,.
\label{2ndscaleint}
\eeq


}


 Contrary to claims in the literature \cite{Solon_Chate, Ikeda}, it is not possible to obtain a
third exact scaling relation. As a result,
we cannot calculate the \emph{correct} exponents for Malthusian flocks.  However, the two scaling laws \rf{1stscaleint} and \rf{2ndscaleint} are sufficient (indeed,  \rf{1stscaleint} by itself is sufficient) to show that the predictions \cite{TT95,Toner_mal} $z=6/5$, $\zeta=3/5$ and $\chi=-1/5$ for the dynamical, anisotropy and roughness exponents are incorrect. 

We will, however, argue that Malthusian flocks {very likely} do display long-range order in two dimensions, and show that fluctuations in some directions of wavevector are smaller than were predicted earlier.

For immortal (i.e., number-conserving) flocks, we show that the proliferation of non-linearities associated with the conserved density field make it impossible to obtain {\it any} exact scaling relations between the scaling exponents. In the absence of such scaling laws, it is impossible to even make an analytic argument that long-ranged order exists in these systems. That they do, 
however, is shown by numerous detailed simulations of both microscopic models whose hydrodynamic behaviour should belong to the universality class of immortal flocks {(see, e.g., \cite{Mahault})}, and continuum equations of the form discussed in this article (see, e.g., Ref.~\cite{chate2020dry}). 
We will further show that several recent articles \cite{Solon_Chate, Ikeda, boost_agnostic} claiming to calculate the exact exponents of flocks are incorrect, {either because they violate certain symmetries, or because they assume symmetries and/or conservation laws that do not, in fact, hold.} 

{Our work deals only with the effect of small fluctuations on the uniformly moving phase of the flock. The {nature of the}  \emph{transition} to that phase is of great interest \cite{chate2020dry, Bertin, Solon_Tailleur_Chate1, Solon_Tailleur_Chate2, Nardini_Tailleur1, Nardini_Tailleur2}, but is not dealt with here. Likewise, we do not examine the stability of flocks to unbinding of topological defects. Thus, our work does not discuss or settle the intriguing recent suggestion \cite{besse2022metastability}, based on numerical simulations, that Malthusian flocks are always destroyed at long enough times by the unbinding of topological defects. There are other examples of nonequilibrium systems in which the unbinding of topological defects generically destroys an ordered state that is stable within a spin-wave theory.
{In two dimensions, the active}  smectic phase \cite{julicher2022broken} is one {such} example.

\section{Equations of motion for Malthusian and Immortal flocks}{\label{EOM}}

We start with the equations of motion for the local polarization ${\bf p}({\bf x},t)$ and number density $\rho({\bf x}, t)$
of a flock, as functions of position ${\bf x}$ and time $t$:
\begin{eqnarray}
&&\partial_{t}
\bp+{\lambda_a (\bp\cdot\vnab)\bp+
\lambda_b (\vnab\cdot\bp)\bp
+\lambda_c\vnab(|\bp|^2)}
 =\nonumber \\&&
U(\rho, |\bp|)\bp -\vnab P_1(\rho, |\bp|) -\bp
\left( \bp \cdot \vnab  P_2 (\rho,|\bp|) \right)\nonumber \\&&+\mu_{B}\vnab
(\vnab
\cdot \bp) + \mu_{T}\nabla^{2}\bp +
\mu_{A}(\bp\cdot\vnab)^{2}\bp+\boldsymbol{\xi}_p \,.
\label{vEOM}
\end{eqnarray}
and
\begin{equation}
	\label{rhoeq}
\partial_t\rho=-\nabla\cdot(\beta {\bf p})+D_1\nabla^2\rho+D_2\nabla\cdot({\bf pp}\cdot\nabla)\rho-h(\rho,|\bp|)+{\xi_n +}\nabla \cdot \boldsymbol{\xi}_\rho \,.
\end{equation}
In equations \rf{vEOM} and \rf{rhoeq}, the coefficients
${\lambda_{a,b,c}}$, 
$U(\rho, |\bp|)$, $h(\rho,|\bp|)$, $\mu_{B,T,A}$,  $D_{1,2}$, {$\beta$}, and the  ``pressures'' $P_{1,2}(\rho,|\bp|)$ are, in general, functions of the flocker number density $\rho$ and the magnitude $|\bp|$ of the local polarization.



We will expand all of them to the order necessary to include all terms that are ``relevant'' in the sense of changing the long-distance behavior of the flock.

The $\bof$, $\bor$, and $\xi_n$  terms are  random, zero-mean,  Gaussian white noises, with correlations:
\begin{eqnarray}
 \la \xi_{p, i}(\br,t)\xi_{p, j}(\br',t') \ra&=&2\Delta_p
\delta_{ij}\delta^{d}(\br-\br')\delta(t-t') \,, \nn\\
\la \xi_{\rho, i}(\br,t)\xi_{\rho, j}(\br',t') \ra&=&2\Delta_\rho
\delta_{ij}\delta^{d}(\br-\br')\delta(t-t') \,, \nn\\
\la \xi_{n}(\br,t)\xi_{n}(\br',t') \ra&=&2\Delta_n
\delta^{d}(\br-\br')\delta(t-t') \,,
\label{white noise}
\end{eqnarray}
where the noise strengths {$\Delta_{p,\rho,n}$} are constant 
{parameters in the coarse-grained theory} (analogous to the temperature in an equilibrium system, as they set the scale of fluctuations), and $i, j$ label
vector components.

These equations are derived purely from symmetry arguments \cite{TT95, TT98, Annals}. However, each term {they contain} has a simple physical interpretation, which we now give.

{We first note that \rf{vEOM} can be viewed as an equation of motion for either a polarization or a velocity, and whether a given term is allowed in the passive limit depends on which of these interpretations we choose. We opt for polarization} in this article.


The $U(\rho, |\bp|)$ term is responsible for spontaneous flock motion. Our analysis will apply to an extremely large class of $U$s; specifically, to all of those that satisfy $U(\rho_0,|\bp|<p_0)>0$, and  $U(\rho_0,|\bp|>p_0)< 0$ in the ordered phase. This last condition {e}nsures that in the absence of fluctuations, the flock will reach a steady state with nonzero polarization magnitude
$|\bp|=p_0$, and $\rho=\rho_0$. The diffusion constants $\mu_{B,T,A}$ reflect the tendency of flockers to {align with} their neighbours. {With the ``polarization'' interpretation for $\bp$, the $P_2(\rho, |\bp|)$ term in (\ref{vEOM}) can arise as the variational derivative of $\int {\bf p}\cdot\nabla P_3$ where $P_3$ is a scalar function of $\rho, |{\bf p}|^2$ \cite{Kung}. The advective $\lambda_a$ term, of course, is non-variational.}

 Both the term $h(\rho,|\bp|)$,
 which embodies birth-death or evaporation-deposition processes, 
and the noise $\xi_n$ (or, equivalently, the noise strength $\Delta_n$) vanish in an ``immortal'' flock, in which, by definition, the number of flockers is conserved. The case in which {both $\xi_{n,\rho}$} are non-zero is called a ``Malthusian'' flock \cite{Toner_mal}. Unsurprisingly, the immortal and Malthusian cases differ considerably, so we will treat them separately below.

For the Malthusian (i.e., non-number-conserving)  dynamics to yield  a stable steady-state density $\rho_0$ in the absence of fluctuations, the birth and death term $h(\rho,|\bp|)$ must satisfy
{$h(\rho_0,p_0) = 0, (\partial_\rho h)_{\rho_0,p_0} > 0$}.

 All of the terms {in \rf{rhoeq}}, with the exception of $h(\rho,|\bp|)$ and the nonconserving part $\xi_n$ of the noise, can be written as the divergence of a current in the immortal case {(in which $h(\rho,|\bp|)$ and $\xi_n$ vanish)}, as is required by number conservation.
Note that this is {\it not} the case for the polarisation equation: many of the terms in that equation are not total divergences, since the polarisation field itself {does not obey a conservation law}. 

We will see in the next two sections that this statement also holds for the Nambu-Goldstone mode of this system; that is, the part of the polarisation field that relaxes slowly {in the ordered phase} for fluctuations at very long wavelengths. 

This is where reference \cite{Solon_Chate} goes wrong:  they claim that the non-stochastic part of the dynamics of the single angle that characterizes the Nambu-Goldstone mode in two dimensions
must be a total divergence and, therefore, must vanish in the limit of small wavenumbers.} 

This is untrue.  The {\it only} requirement of a Nambu-Goldstone mode {- which is the small amplitude, small wavenumber fluctuation of an order parameter corresponding to a broken continuous symmetry (see, for instance, \cite{Low_Manohar}) -} is that it relaxes slowly
in the limit of long wavelengths {with the relaxation rate vanishing in the limit of infinite wavelength in which case the mode corresponds to the global transformation of the order parameter under the broken symmetry}; this by no means implies that all terms in the equation of motion for a Nambu-Goldstone mode must be expressible as total divergences. Indeed, numerous counter-examples - that is, systems whose Nambu-Goldstone modes have terms in their equations of motion that can {\it not} be written as total derivatives -  exist: the KPZ equation \cite{KPZ}\footnote{{If the field $H$ of the KPZ equation is a non-compact field, then it represents the dynamics of the Nambu-Goldstone mode of broken $D$-dimensional spatial translation symmetry by a $D-1$-dimensional membrane \cite{Low_Manohar}. If it is a compact field, it can represent the Nambu-Goldstone mode of driven or active periodic systems \cite{Aranson_ref, Saha_ref} or a variety of time-crystalline systems, that break parity, ranging from active XY models \cite{Basu1, Basu2} to chiral active liquid crystals \cite{AM_hex, AM_chi_rev} to non-reciprocal systems \cite{Fruchart_nonreci} to driven-dissipative condensates \cite{compKPZ1, compKPZ2, compKPZToner, Diehl_new, Diehl_rev}.}
} 
for one, and the hydrodynamics of {nematic fluids \cite{Stark}}, crystalline solids \cite{mpp}, smectics \cite{mpp}, and discotic phases \cite{discodyn} for {a few} others. {In fact, except in very special circumstances, \emph{all} active and passive liquid-crystalline and crystalline states serve as counterexamples.} We will now briefly describe how each of {the} systems {we named} is a counter-example to the claim that time derivatives of Nambu-Goldstone modes must be total divergences.

The nonlinearity in the KPZ equation 
\beqn
\pp_t H=\nu\nabla^2 H+{\frac{\lambda}{2}}(\nabla H)^2+f\,,\label{KPZ}
\eeqn
i.e., the $\lambda$ term, can clearly {\it not} be written as a total derivative. This means that, if the assertion of \cite{Solon_Chate} were correct, that $\lambda$ term would be forbidden{, which it is not}. 

Note further in this context that, if the $\lambda$ term in \rf{KPZ} {\it was}, somehow, expressible as a total divergence, then one would be inevitably led to the conclusion that the noise strength - i.e., the correlator for the noise $f$ in \rf{KPZ}- would be unrenormalized. In fact, it has been known \cite{KPZ, FNS} since the KPZ equation was first proposed that the noise strength is, indeed, renormalized.

Likewise, in the immortal flocking problem we study here, the noises will also be renormalized. This has been explicitly demonstrated to one loop order for $d>2$ Malthusian flocks \cite{Mal_RG1, Mal_RG2}, and will certainly hold for immortal flocks as well, contrary to the claims in \cite{Solon_Chate}, which arise entirely from their erroneous assertion that the time derivative of any Nambu-Goldstone mode must be a total divergence.

Three of the other counter-examples we quoted involve translationally ordered systems, for which the Nambu-Goldstone modes are well known \cite{mpp} to be a displacement field $\bu(\br,t)$, where $\bu$ has $3$ components
in a crystalline solid, $2$ components in a ``discotic'', and one component in a ``smectic''. In all three examples, the equation of motion for the displacement field is
\beq
\pp_t\bu=\bv_\parallel+...
\label{cryseom}
\eeq
where $\bv_\parallel(\br,t)$ is the projection of the local velocity field onto the directions of broken translational symmetry, and the ellipsis denotes terms involving spatial derivatives. Here the fact that the time derivative of the Nambu-Goldstone mode is not a total divergence is even more obvious: the $\bv_\parallel(\br,t)$ term doesn't even involve spatial
derivatives, much less total {divergences.} {Similarly, the dynamics of the Nambu-Goldstone mode in active or driven solids \cite{AM_sol, Rangan} or polar active smectics \cite{Chen_Toner} are not expressible as total divergences.}

{
{Finally, for states with spontaneously broken spatial rotation symmetry, the Nambu-Goldstone mode is an angle field $\theta$ (for the simplest case of two space dimensions), and it is advected by the velocity field:}
\begin{equation}
\partial_t\theta+{\bf v}\cdot\nabla\theta=...
\end{equation}
where the ellipsis contains other couplings to the velocity field as well as relaxational and stochastic terms. Importantly, ${\bf v}\cdot\nabla\theta$ is \emph{not} a total derivative and a term $\theta\nabla\cdot{\bf v}$ whose presence could have converted this to a total derivative is forbidden because it is not consistent with rotation symmetry ({$\theta\nabla\cdot{\bf v}$ transforms the same way as $\theta$ under rotation and a term $\propto\theta$ is clearly forbidden by rotation symmetry in the equation for $\partial_t\theta$}; see {also} the symmetry transformation \eqref{rotinv} in the next section)\footnote{{Similarly, and contrary to a claim by \cite{boost_agnostic}, a term $\propto\theta(\nabla\cdot{\bf p})$ is forbidden in the angular dynamics of polar flocks that we will discuss in the next sections.}}. 
A similar argument can be constructed for polar and nematic liquid crystals in higher dimensions as well\footnote{{In fact, even the Nambu-Goldstone mode equation that can be derived from the equilibrium model A dynamics governed by 
a free energy 
functional with a term $\propto \int {\bf p} \cdot \nabla P_3(\rho)$ 
cannot be written as a total divergence. In two dimensions, the angular dynamics has the form of \eqref{thetaeq}, which we will discuss in Sec. \ref{eomimmort}, with $\lambda=g_1=0$. {Further, even the functional derivative of the two Frank-constant free energy of a two-dimensional nematic liquid crystal \cite{Nelson_Pelcovits} with respect to $\theta$ cannot be written as a total divergence. Of course, the dynamics of $\theta$ in active nematics can also not be written as a total divergence \cite{Shradha_aditi, Suraj_PRE}.}}}.
}

So the claim of \cite{Solon_Chate} that the time derivative of any Nambu-Goldstone mode must be a total divergence is not only unsupported by any argument, but is in fact explicitly contradicted by {some} of the most well-known examples of Nambu-Goldstone {modes.}

The correct equations of motion for flocks are our equations \rf{vEOM} and \rf{rhoeq}, which lead, for immortal flocks, to an equation of motion for the Nambu-Goldstone mode that {\it does} contain non-total divergence terms, as we'll see in the  section\rf{Immortal}.

The rate of change of the Nambu-Goldstone mode in Malthusian flocks {\it does} prove to be writable as a total divergence. However, this is a derived result, not a consequence of imposing conservation of the Nambu-Goldstone mode at the outset.
Rather, it proves to be a consequence of rotation invariance, as we'll show in section \rf{malt}{, which, however, does not apply to immortal flocks}.

We'll now describe the long-wavelength theories of our two cases: Malthusian, and immortal, starting with the simpler Malthusian case.


\section{Hydrodynamics of Malthusian flocks}{\label{malt}}

\subsection{Equation of motion}{\label{eommalt}}

We are interested in the long-time, large-scale dynamics of the ordered phase of a Malthusian flock.   We will begin by considering the simple limit in which the {turnover rate {$\tau_\rho^{-1}=(\partial_\rho h)_{\rho_0,p_0}$} is so large}  that the density is locked on to that value, and never changes. Later, we will argue that relaxing this constraint changes nothing in the long-wavelength limit. That is, in the language of the renormalization group, departures of the {turnover} rate from {infinitely fast relaxation} are ``irrelevant''. 

In the absence of noise and density fluctuations, this ordered state will
simply be any one of the infinity of uniform steady states of the equations of motion \rf{vEOM} and \rf{rhoeq}
\beq
\bp(\br,t)=p_0 \hn \,,
\label{ss}
\eeq
where $p_0$ is the value of the magnitude of the polarization at which $U(\rho_0, |\bp|)$ vanishes. Here, as a consequence of the underlying rotation invariance of our dynamics, the direction $\hn$ of the polarization in the steady state is completely arbitrary (which is why there is an infinity of such states).

To study fluctuations about this uniform state,
we write the polarization as
\beq
{\bf p(\br,t)}=(p_0+\delta p(\br,t))(\cos\theta(\br,t),\sin\theta(\br,t)) \,,
\label{pexpmalt}
\eeq
where we have chosen our coordinate system with its $x$-axis pointing along the direction $\hn$ of the spontaneously broken symmetry.

Inserting \rf{pexpmalt} into the equation of motion  for the polarization, and expanding to linear order in $\dep$, gives
\bew
\beq
(\cos\theta, \sin\theta)\pp_t\delta p+{p_0}(-\sin\theta, \cos\theta)\pp_t\theta=-\alpha\delta p(\cos\theta, \sin\theta)+... \,,
\label{pthetaeom}
\eeq
\ew
where we've defined $\alpha\equiv{ -p_0}\left({\pp U(\rho, |\bp|)\over\pp |\bp|}\right)_{\rho_0, p_0}${$>0$}, and the ellipsis denotes terms involving spatial derivatives of $\theta$,  and
$\delta p$, as well as products of $\delta p$ and spatial derivatives of $\theta$. These terms all vanish in the limit of very small spatial gradients.

Note that the left hand side of this equation \rf{pthetaeom} neatly separates into orthogonal components along
the unit vector ${\hn}=(\cos\theta, \sin\theta)$ along $\bp$ and the vector ${\hn_\perp=}(-\sin\theta, \cos\theta)$ perpendicular to it. 
Therefore, projecting  \rf{pthetaeom} along ${\hn}$ by taking the dot products of both sides of  \rf{pthetaeom} with ${\hn}$ leads immediately to an equation of motion for $\dep$:
\beq
\partial_t\delta p=-\alpha\delta p+... \,,
\label{dpeom}
\eeq
where once again the ellipsis denotes terms that vanish in the limit of very small spatial gradients.

We see immediately from \rf{dpeom} that $\delta p$ is a ``fast'' variable, in the sense that it relaxes with a finite lifetime $\tau=|\alpha|^{-1}$ to a value determined by the ellipsis in \rf{dpeom}, {similarly to the density fluctuations} {with the relaxation timescale $\tau_\rho$}.  Note also that $\pp_t \delta p$ is negligible compared to the $|\alpha|\delta p$ term since we're considering very slow modes. Hence, that  $\pp_t \delta p$ term can be dropped from \rf{dpeom}. Once we do so, \rf{dpeom} becomes a simple linear equation that relates
$\dep(\br, t)$ to the {\it instantaneous} value of $\theta(\br, t)$ and its spatial derivatives.

A further simplification follows from recognizing that, since the ellipsis $...$  in \rf{dpeom} {always} 
involves spatial derivatives of $\theta$, and we are interested in situations in which all spatial derivatives are small, because we are investigating the long-wavelength (i.e., hydrodynamic) limit, it follows that $\delta p$ itself is small, and so can be dropped from the ellipsis (where it always either multiplies gradients of $\theta$, or is differentiated itself).

Thus, the end result of this analysis of the equation of motion \rf{dpeom} for $\dep$ is that
$\dep$ can be expressed entirely in terms of the local instantaneous value of $\theta$ and {its} spatial derivatives.


We now insert this solution for $\dep$ into the equation of motion for $\theta$. The former can be obtained by looking at the component of the equation of motion \rf{pthetaeom} along the vector {$\hn_\perp$} perpendicular to the polarization, as can be done by taking the dot products of both sides of \rf{pthetaeom} with {$\hn_\perp$}. Having eliminated $\dep$ as described above, this equation can only involve $\theta$ and its derivatives.

Rather than explicitly go through the algebra just described, we will instead simply note that the net result of such an analysis must be a closed set of equations of motion for $\theta$  that respect the symmetries and conservation laws of our system. {In two dimensions, the symmetries are reflection about the $x$-axis (that is, $y\to-y${, $\theta\to-\theta$}) and \emph{spatial} rotation by an arbitrary angle, i.e., 
\begin{widetext}
\begin{equation} \label{rotinv}
\theta (\br,t)\to\theta(\br,t)+\psi, \, x\to x\cos\psi-y\sin\psi, \, y\to y\cos\psi+x\sin\psi \,,
\end{equation}
\end{widetext}
where $\psi$ is any constant rotation angle.} 
Note that the dynamical equation for $\theta$ obtained by \cite{Ikeda} does not possess the symmetries described here - specifically,  rotational symmetry \rf{rotinv} {(see, for instance, their Eq. 16)} - and is therefore incorrect\footnote{A recently published erratum \cite{Err_Ikeda}acknowledges that the analysis in \cite{Ikeda} was incorrect.}.

The complete \footnote{{We retain all relevant terms. This includes nonlinearities with one factor of $\nabla$; nonlinearities with two factors of $\nabla$ will be shown to be irrelevant.}} hydrodynamic equation of motion for $\theta$ for Malthusian flocks is
\begin{equation}
	\label{thetaeqmal}
	\partial_t\theta=\lambda\nabla\cdot(-\sin\theta,\cos\theta)+D_x\partial_x^2\theta+D_y\partial_y^2\theta+\xi_\theta \,,
\end{equation}
where  the noise $\xi_\theta$ is Gaussian and zero-mean, with two-point correlations
\beq
\langle{\xi}_\theta({\bf x}, t){\xi}_\theta({\bf x}', t`)\rangle=2{\Delta}_\theta\delta({\bf x}-{\bf x}')\delta(t-t') \,.
\label{thetanoisemalt}
\eeq
 We have verified that systematically eliminating $\dep$ {as} described above recovers this form. This approach also gives relations between the parameters $\lambda$, $D_x$, $D_y$, and $\Delta_\theta$ and the parameters of our original equation of motion \rf{vEOM} and \rf{rhoeq}. We will not bother giving those relations here, since all of the parameters of \rf{vEOM} and \rf{rhoeq} are phenomenological in any case (that is, {they are not calculated but, in principle, can be obtained from a fit to experiment),}
and one could just as well directly fit $\lambda$, $D_x$, $D_y$, and $\Delta_\theta$ to experiment instead.

This equation of motion for 2{D} Malthusian flocks was also obtained by \cite{Solon_Chate}.

\subsection{Linear theory}{\label{maltlin}}

If we neglect the part of the $\lambda$ term that is nonlinear in $\theta$ in \rf{thetaeqmal}, that equation becomes
\beq
	\label{thetaeqmallin}
	\partial_t\theta=-\lambda\pp_x\theta+D_x\partial_x^2\theta+D_y\partial_y^2\theta+\xi_\theta \,.
\eeq
We can eliminate the $\lambda\pp_x\theta$ term in
\rf{thetaeqmal}
by
{changing co-ordinates to 
\beq
\br'\equiv\br-\lambda t  {\bf \hat x}
\label{boost}
\eeq
in a new Galilean frame moving with respect to our original frame
in the direction ${\bf \hat{x}}$ of mean flock motion at speed $\lambda$.}
This changes \rf{thetaeqmallin} to
\beq
	\label{thetaeqmalb}
	\partial_t\theta=D_x\partial_x^2\theta+D_y\partial_y^2\theta+\xi_\theta \,,
\eeq
where we have dropped the primes on {the transformed coordinates.} 

We can directly read off the scaling exponents implied by the linear theory from \eqref{thetaeqmalb}. Since the $D_x$ and $D_y$ terms contain the same number of derivatives, if we define an ``anisotropy exponent'' $\zeta$ by saying that characteristic distances $L_x$ along $x$ scale with distances $L_y$ along $y$ according to
\beq
L_x\propto L_y^\zeta \,,
\label{zetadef}
\eeq
we clearly have
\beq
\zeta=\zeta_{\rm lin}=1 \,;
\label{zetalin}
\eeq
that is, the scaling is (according to this linear theory) isotropic.

We can also define a ``dynamic'' exponent $z$ by the scaling of characteristic times
$\tau$ with $L_y$ via
\beq
\tau\propto L_y^z \,.
\label{zdef}
\eeq
 Since \eqref{thetaeqmalb} says that one time derivative is proportional to two $y$ derivatives, we must have
\beq
z=z_{\rm lin}=2 \,.
\label{zlin}
\eeq 
{From \rf{thetaeqmallin} and \rf{thetanoisemalt} it is straightforward to show that the spatial Fourier transform of the equal-time correlator $C^{\theta}({\bf r}) \equiv \langle \theta({\bf 0},t) \theta({\bf r},t) \rangle$ is
\begin{equation}
  C^{\theta}_{\bf q}  =\frac{\Delta_\theta}{q^2(D_x\cos^2\phi+D_y\sin^2\phi)} \,,
	\label{cqet}
\end{equation}
}
where $\phi$ is the angle between the wavevector $\bq$ and the $x$-axis {and $q\equiv|{\bf q}|$}. 

From this, we see that the three parameters $\Delta_\theta$, $D_x$, and $D_y$ completely determine the scale of the $\theta$-fluctuations in the linear theory. (Of course, this is hardly surprising, since they are the {\it only} parameters of the linear theory, except for irrelevant terms that we've neglected.)  We will use this fact in the Dynamical Renormalization Group (DRG) of the next subsection to guide us to the most convenient choice of some otherwise arbitrary rescaling exponents.

Integrating \rf{cqet} overall $\bq$ gives the real-space fluctuations of $\theta$:
\beq
	\langle\theta^2(\br,t)\rangle=\frac{\Delta_\theta}{2\pi\sqrt{D_xD_y}} \ln(\Lambda L) \,,
	\label{cret}
\eeq
where we've introduced an ultraviolet cutoff $q_{\rm max}=\Lambda$ and an infrared cutoff $q_{\rm min}={1\over L}$, where $L$ is the system's linear spatial extent. {Typically, the ``roughness'' exponent $\chi$ is defined as,}
\beq
\theta_{\rm typical}\sim L_y^\chi \,
\label{chidef}
\eeq 
{which, for the logarithmic case of \eqref{cret}, corresponds to}
\beq
\chi=\chi_{\rm lin}=0 \,.
\label{chilin}
\eeq

We will next use the DRG to show that these linear results are radically changed by the $\lambda$ nonlinearity in \eqref{thetaeqmalb}.

\subsection{Dynamical Renormalization Group (DRG)  analysis} {\label{rgmalt}}

Our approach is precisely that of \cite{FNS}.

First we decompose the Fourier modes $\theta(\bq, \omega)$  into  a rapidly varying part
$\theta^>(\bq, \omega)$   and  a slowly varying part $\theta^<(\bq, \omega)$  in the equation of motion  \rf{thetaeqmal}. The rapidly varying part is supported in the {cylindrical} momentum shell 
$-\infty <q_x<\infty$, $\Lambda b^{-1}<|q_y|<\Lambda$, where $\Lambda$ is the ultraviolet cutoff, and $b$ is an arbitrary rescaling factor that we will ultimately take to be $b=1+d\ell$, with $d\ell$ {infinitesimal}.  The slowly varying part is supported in $-\infty <q_x<\infty$, $0<|q_y|<b^{-1}\Lambda$. We separate the noise $\xi_\theta$ in exactly the same way.

The  DRG procedure then consists of two steps. In step 1, we eliminate $\theta^>(\bq, \omega)$  from  \rf{thetaeqmal}. We do this by solving iteratively for $\theta^<(\bq, \omega)$. This solution is a perturbative expansion in $\theta^>(\bq, \omega)$, as well as the fast components $\xi_\theta^>$ of the noise. As usual, the perturbation theory can be represented by Feynman graphs. We substitute these solutions into  \rf{thetaeqmal}  and average over the short wavelength components $\xi_\theta^>(\bq, \omega)$ of the noise $\xi_\theta$, which gives a closed EOM  for $\theta^<(\bq, \omega)$.

In step 2, we  rescale the {\it real space} field $\theta^<(\br, t)$, time $t$, and coordinates $x$ and $y$ as follows:
 \beq
y\to b y \sep t\to b^z t \sep x\to b^\zeta x \sep  \theta\to b^\chi\theta \,.
\label{rescale}
\eeq

At this point, the choice of the rescaling exponents $z$, $\zeta$, and $\chi$ is arbitrary. However, standard DRG arguments \cite{FNS} show that
the values of these exponents that produce stable fixed points of the RG process are also the values of $\zeta$, $z$, and $\chi$ defined by the scaling laws \rf{zetadef}, \rf{zdef}, and \rf{chidef}.

 After these two RG steps, we reorganize the resultant EOMs so that they have the same form as \rf{thetaeqmal}, but with various coefficients renormalized. Note that the renormalized equation {\it must} have the same {symmetries} as the original \rf{thetaeqmal}. 


 This reorganization amounts to multiplying the EOM by a power of $b$ chosen to restore the coefficient of $\pp_t \theta$ to unity. 

The RG now proceeds by iterating this process.  

It is useful in performing the DRG to expand the equation of motion \rf{thetaeqmal} in powers of $\theta$. This is easily seen to lead, after the boost \rf{boost}, to
\beq
	\label{thetaeqmalexp}
	\partial_t\theta=\sum_{n=1}^\infty a_n\theta^{2n}\pp_x\theta+\sum_{n=0}^\infty c_n\theta^{2n+1}\pp_y\theta+D_x\partial_x^2\theta+D_y\partial_y^2\theta+\xi_\theta \,,
\eeq
where the expansion coefficients before we begin renormalizing (i.e., on DRG step $m=0$, where $m$ counts the number of iterations of the DRG) are given by
\beq
a_n(m=0)=-{(-1)^n\lambda\over(2n)!} \sep c_n(m=0)=-{(-1)^n\lambda\over(2n+1)!} \,.
\label{bareexp}
\eeq
These expressions will not continue to hold as we iterate the DRG, although it is important to note that{, because of rotation invariance,} it is only the rescaling step of the DRG that changes the ratios $a_n/a_{n^\prime}$,  $c_n/c_{n^\prime}$, and, most importantly, $c_n/a_{n^\prime}$. {That is, the ``intermediate'' coefficients $a_n^I$ and $c_n^I$ obtained \emph{after} averaging out the short wavelength modes but \emph{before} rescaling must satisfy the equation $a_n^I/c_n^I=a_n/c_n$.} 
Note that there would have been an $n=0$ term in the first ($a_n$) sum had we not done the boost \rf{boost}, which eliminated it.

The recursion relations between the parameters of \rf{thetaeqmalexp} on successive steps of the RG are
\beqn
 a_n^\prime&=& b^{2n\chi+z-\zeta} (a_n + {\rm graphs}) \,, \label{anres}\\
 c_n^\prime&=& b^{(2n+1)\chi+z-1} (c_n + {\rm graphs}) \,, \label{cnres}\\
 D_x^\prime&=&b^{z-2\zeta}(D_x+ {\rm graphs}) \,, \label{dxres}\\
 D_y^\prime&=&b^{z-2}(D_y+ {\rm graphs}) \,, \label{dyres}\\
 \Delta_\theta^\prime&=&b^{z-\zeta-2\chi-1}(\Delta_\theta+ {\rm graphs}) \,,\label{delthetares}
\eeqn
where the primes denote parameters after the RG step in question, and ``graphs'' denote the renormalizations arising from averaging over the short-wavelength degrees of freedom.

Even without evaluating the graphical corrections in these expressions \rf{anres}-\rf{delthetares}, we can use them to determine which of the nonlinearities $a_n$ and $c_n$ are ``relevant'', in the RG sense of changing the behaviour of the system from that predicted by the linear theory described earlier. We can do so by first noting that, in the absence of those nonlinearities, the graphical corrections must vanish, since the nonlinearities are the only terms in the equations of motion that couple modes at different wavevectors, and, hence, the only couplings between the ``fast'' and ``slow'' modes. Hence, if we choose the rescaling exponents $z$, $\zeta$, and $\chi$ in \rf{dxres}-\rf{delthetares} to keep the linear terms $D_{x,y}$ and the noise strength $\Delta_\theta$ fixed (so that the scale of the fluctuations remains the same upon renormalization), then we can assess the relevance or irrelevance of the nonlinearities $a_n$ and $c_n$ when those nonlinearities are initially small (so that the graphical corrections can be ignored in the early RG steps)  by asking simply whether or not they grow upon renormalization if they start out being small.

Ignoring the graphical corrections, we see that equations \rf{dxres}, \rf{dyres}, and \rf{delthetares} imply that, to keep $D_{x,y}$ and the noise strength
$\Delta_\theta$ fixed, we must choose the rescaling exponents $z$, $\zeta$, and $\chi$ so that the following three conditions are satisfied:
\beq
z-2=0 \sep z-2\zeta=0 \sep z-\zeta-2\chi-1=0 \,,
\label{lincond}
\eeq
whose solutions are trivially found to be 
\beq
z=2 \sep \zeta=1 \sep \chi=0 \,,
\label{zlinrg}
\eeq
which the alert reader will recognize as the exponents we found earlier by inspecting the correlation function predicted by the linear theory.

Using these exponents in \rf{anres} and \rf{cnres}, we get
\beqn
    a_n^\prime&= b(a_n + {\rm graphs}) \,, \label{anlinexp}\\
        c_n^\prime&= b (c_n + {\rm graphs}) \,. \label{cnlinexp}
       \eeqn
We see from this immediately that {\it all} of the
nonlinearities $a_n$ and $c_n$ are {\it relevant} (that is, grow upon renormalization) if they are initially small (so that the ``graphs'' can be neglected). This proves that the linear results (including \rf{zlinrg}), are {\it never} valid in a two dimensional Malthusian flock, except in the very special case in which $\lambda=0$, which will never occur in practice.

The long wavelength physics will therefore always {differ from that of the linear theory,} 
and must be controlled by an RG fixed point at which at least one of the nonlinearities  $a_n$
and $c_n$ are nonzero.

We'll now argue that only two of the infinity of nonlinearities in \rf{thetaeqmalexp} actually need to be considered; namely, $a_1$ and $c_1$. To see this, we begin by arguing that the graphical corrections to $\Delta_\theta$ are $0$. This follows from the fact that all of the nonlinear terms in \rf{thetaeqmalexp} are total divergences. Hence, they can only lead to renormalizations of the noise that are total divergences,  while the noise  $\xi_\theta$ is not. Hence, the contribution of those nonlinearities to the Fourier transformed noise correlation $\langle| \xi(\bq, \omega)|^2\rangle$  must be proportional to at least two powers of $\bq$. Since $\Delta_\theta$ is independent of $\bq$, it therefore cannot be renormalized by any of the nonlinear terms. Hence the ``graphs'' in equation \rf{delthetares} must vanish.

Therefore, to obtain a fixed point, equation \rf{delthetares} implies that we must have
\beq
z-\zeta-2\chi-1=0
\label{delthetafix}
\eeq
even at any {\it non}linear fixed point.
A $4-\epsilon$ expansion for Malthusian flocks \cite{Mal_RG1, Mal_RG2} finds that, near four spatial dimensions, the correction to the diffusion constants $D_{x,y}$ are all positive. It seems plausible to assume that this remains true all the way down to $d=2$.

This implies that, to keep $D_y$ fixed, we must have
\beq
z-2<0 \,,
\label{dyfix}
\eeq
while to keep $D_x$ fixed, we must have
\begin{equation}
z-2\zeta <0 \,.
\label{dxfix}
\end{equation}
This last condition clearly implies
\beq
{z\over2}<\zeta \,,
\label{dxfix2}
\eeq
while \rf{dyfix} implies
\beq
z<2 \,.
\label{dyfix2}
\eeq

Together with \rf{delthetafix},  \rf{dxfix2} implies
\beq
\chi<{z\over4}-{1\over2} \,.
\label{chiineq1}
\eeq
 Using \rf{dyfix2} in this implies
 \beq
 \chi<0 \,.
 \label{chiineq}
 \eeq
This is an extremely important result, since it implies that {the long-range ordered state survives fluctuations}. That is, it says that the characteristic fluctuations of $\theta(\br,t)$ in real space, {whose $L{_y}$-dependent part scales} like $L{_y}^\chi$, remain finite as $L{_y}\to\infty$. 

 For the purposes of identifying the relevant nonlinearities in \rf{thetaeqmalexp}, the negativity of $\chi$ implies, as can
 be immediately seen from
\rf{anres} and \rf{cnres}, that the most relevant terms of the set $a_n$, $c_n$ in  \rf{thetaeqmalexp} are those with the smallest value of $n$ since the graphical corrections are the same for all $n$ as required by rotation invariance; that is, {$n=1$ for $a_n$ and $n=0$ for $c_n$ (we can't eliminate $a_1$ using this argument even though its recursion relation has a higher power of $\chi$ than the recursion relation for $c_0$ because we don't know the value that $\zeta$ will assume at the fixed point, where we have argued $\chi<0$)}.

Therefore, the minimal theory that  predicts the large-distance, long-time behaviour of such flocks is of the form
\begin{equation}
	\label{angnl2}
	\partial_t\theta={\lambda}_1\partial_y\left(\frac{\theta^2}{2}\right)+{\lambda}_2\partial_x\left(\frac{\theta^3}{6}\right)+D_x\partial_x^2\theta+D_y\partial_y^2\theta+\xi_\theta,
	\end{equation}
with, before renormalization, $\lambda_1=-\lambda=-\lambda_2$.

This is the model we will now analyse using the DRG.

We begin by rewriting the recursion relations in the form of differential equations. This very standard approach proceeds as follows:
as already mentioned, we choose $b=1+d\ell$ with $d\ell$ differential. Instead of keeping track of the number $n$ of iterations of the renormalization group, we introduce a ``renormalization group time'' $\ell$ defined as $\ell\equiv nd\ell$. This renormalization group ``time'' should {\it not} be confused with the actual time $t$ in our problem; the two are completely different! And, instead of speaking of the values of the parameters after $n$ iterations, we think of them as continuous functions of the continuous variable $\ell$.

Taking the difference between successive values of the parameters and dividing by $d\ell$ leads to differential recursion relations for the parameters, which, based entirely on equations \rf{anres}-\rf{delthetares}, and our earlier argument that there are no graphical corrections to
$\Delta_\theta$, can be written
\begin{subequations}
\begin{equation}
 {dD_x\over d\ell}=\bigg[z-2\zeta+f_{_{D_x}}(\lambda_1, \lambda_2, D_x, D_y, \Delta_\theta)\bigg]D_x \,,
\label{dxrr} 
\end{equation}
\begin{equation}
{dD_y\over d\ell}=\bigg[z-2+f_{_{D_y}}(\lambda_1, \lambda_2, D_x, D_y, \Delta_\theta)\bigg]D_y \,,
\label{dyrr} 
\end{equation}
\begin{equation}
{d\Delta_\theta\over d\ell}=\bigg(z-\zeta-2\chi-1\bigg)\Delta_\theta \,,
\label{dthetrr} 
\end{equation}
\begin{equation}
{d\lambda_1\over d\ell}=\bigg[\chi+z-1+f_\lambda(\lambda_1, \lambda_2, D_x, D_y, \Delta_\theta)\bigg]\lambda_1 \,,
\label{lambda1rr} 
\end{equation}
\begin{equation}
{d\lambda_2\over d\ell}=\bigg[2\chi+z-\zeta+f_\lambda(\lambda_1, \lambda_2, D_x, D_y, \Delta_\theta)\bigg]\lambda_2 \,,
\label{lambda2rr}
\end{equation}
\end{subequations}
where the $f_{_{D_x}}$ etc denote the ``graphical corrections''.

Two features of these recursion relations require some explanation:

\noindent 1) the recursion relation \rf{dthetrr} for $\Delta_\theta$ is {\it exact}, since, as shown earlier, there are no graphical corrections to $\Delta_\theta$. As we discussed earlier, this implies the exact scaling relation \rf{delthetafix}.

\noindent 2) The graphical corrections $f_\lambda$ to $\lambda_1$ and $\lambda_2$ are equal. This follows from the underlying rotation invariance of the dynamics. Specifically, the original form of the nonlinearity $\nabla\cdot(-\sin\theta,\cos\theta)$ is fixed by rotation invariance and implies a relation amongst the coefficients of the powers of $\theta$ upon expanding the trigonometric functions. That is, if we write
$\lambda_1\theta\partial_y\theta\equiv-\lambda\theta\partial_y\theta$, $\lambda_2\theta^2\partial_x\theta\equiv(\lambda\theta^2/2)\partial_x\theta$, the ratio between $\lambda_1$ {and} $\lambda_2$ must be preserved under the perturbative step of the RG. Of course, these ratios can and will trivially change on the rescaling step.

Now consider in light of this observation what happens when we perform the first part of this RG step, in which we integrate out large-wavenumber modes. The ``intermediate'' values of
$\lambda_i^I(\ell+d\ell)$,  where $i=1,2$, after the integration part of the RG step that takes us from RG ``time'' $\ell$ to $\ell+d\ell$,  but before the rescaling part of this step, can be written in complete generality
\beq
\lambda_i^I(\ell+d\ell)=\lambda_i(\ell+d\ell)[1+f_i(\lambda_1, \lambda_2, D_x, D_y, \Delta_\theta)d\ell] \,.
\label{lamint}
\eeq
But since the ratio of $\lambda_1$ to $\lambda_2$ is  be preserved on this step, we must have
\beq
f_1(\lambda_1, \lambda_2, D_x, D_y, \Delta_\theta)
=f_2(\lambda_1, \lambda_2, D_x, D_y, \Delta_\theta) \,.
\label{lambdaexact}
\eeq
Performing the rescaling part of the RG and constructing the differential recursion relations as described above then immediately implies
\rf{lambda1rr} and \rf{lambda2rr}, with $f_\lambda$ the common value of $f_1$ and $f_2$.

We can argue that $f_\lambda$ must vanish if $\lambda_2$ vanishes. {To show this, we begin by noting that when $\lambda_2=0$,} Eq. \eqref{angnl2} has a ``pseudo-Galilean invariance'' under the transformation $\theta\to\theta+\theta_0$, $y\to y+\lambda_1\theta_0t$, with $\theta_0$ being an arbitrary constant. {This invariance} must continue to hold,  {\it with the same value of $\lambda_1$}, under renormalisation
\cite{Toner_mal, Toner_rean, TT98}. This implies that $f_\lambda$ must vanish {when $\lambda_2=0$. Analyticity in both $\lambda$`s then implies} that $f_\lambda$ {can be no larger than} order $\lambda_2$ for small $\lambda_2$.

Now let us consider the possible types of fixed points of these recursion relations.

Since we have already shown that a fixed point with all of the nonlinearities
equal to zero is always unstable under renormalization, only three possibilities for the values $(\lambda_1^*, \lambda_2^*)$ of $\lambda_{1,2}$ at a stable fixed point remain:

\noindent i) $\lambda_1^*\ne0$, $\lambda_2^*=0$.

\noindent ii) $\lambda_1^*=0$, $\lambda_2^*\ne0$.

\noindent iii) $\lambda_1^*\ne0$, $\lambda_2^*\ne0$.

We will now show that only possibility (iii) is, in fact, possible. We'll show this by demonstrating that fixed points of the type (i) and (ii) are unstable under renormalization.

Consider the first possibility (i), which is the type of fixed point investigated by Toner \cite{Toner_mal} (who, alas, failed to recognize that it was unstable). At such a fixed point, the equation of motion \rf{angnl2} is identical to a model for a completely unrelated physical system studied by Hwa and Kardar \cite{Hwa}.\footnote{{{Hwa and Kardar`s} model was \emph{not} rotation invariant to start with and, therefore, the considerations that forced us to retain $\lambda_2$ didn't apply.}}.  Any such fixed point must be unstable against $\lambda_2$. To see this, note that, at such a fixed point, if we linearize the recursion relation \rf{lambda2rr} for $\lambda_2$, the $f_\lambda$ term drops out. This is because, as noted above, $f_\lambda$ is at least of order
$\lambda_2$ at such a fixed point. Hence, its contribution to the recursion relation for
$\lambda_2$ at a fixed point of the type (i) is of $O(\lambda_2^2)$ at most, and hence vanishes upon linearization.

This leaves the recursion relation for 
 $\lambda_2$ as
\beq
{d\lambda_2\over d\ell}=\left( 2\chi+z-\zeta\right)\lambda_2 \,.
\label{l2fp1rr}
\eeq

At a fixed point of the type (i), it is possible to actually determine the exponents exactly. One can do this by noting that, {as we argued earlier,}  $f_\lambda$ vanishes at such a fixed point due to pseudo-Galilean invariance, since $\lambda_2=0$ there. Equation \rf{lambda1rr} therefore implies that, to keep $\lambda_1$ fixed and non-zero (which, by assumption, it is at this fixed point), we must have
\beq
\chi+z-1=0 \,.
\label{l1fixfp1}
\eeq

In addition, when $\lambda_2$ is zero, as we are assuming for type (i) fixed points, $D_x$ cannot renormalize, since the only nonlinearity is a total $y$-derivative, which cannot generate a term like the $D_x$ in \rf{thetaeqmalexp}, which is a total $x$ derivative. Hence, the graphical correction $f_{D_x}$ in the recursion relation \rf{dxrr} for $D_x$ must vanish at the fixed point (i). Therefore, to keep $D_x$ fixed, \rf{dxrr} implies
\beq
z=2\zeta \,.
\label{dxfix1}
\eeq

Note that we now have three linearly independent equations (\rf{l1fixfp1}, \rf{dxfix1}, and \rf{delthetafix}) for the three exponents $\chi$, $\zeta$, and $z$. Solving these gives
\beq
\chi=-{1\over5} \sep \zeta={3\over5} \sep z={6\over5} \,,
\label{canon}
\eeq
which are the exponents obtained by Toner \cite{Toner_mal}. 

Unfortunately, as can be immediately seen from the linearized recursion relation (\ref{l2fp1rr})  for $\lambda_2$ at a fixed point of the type (i), these exponents imply that such a fixed point is unstable against $\lambda_2$. Indeed, plugging the exponents\rf{canon}
into 
(\ref{l2fp1rr}) gives
\beq
{d\lambda_2\over d\ell}={1\over5}\lambda_2 \,,
\label{l2fp1rr2}
\eeq
which clearly shows that this fixed point is unstable. 

We note in passing 
that in an ``easy-plane'' Malthusian flock in three dimensions \cite{toner2024birth}, which has a very similar structure to the two-dimensional Malthusian flock we study here, an analysis like that just presented implies that $\lambda_2$ is actually {\it irrelevant} at a fixed point of type (i), which is therefore stable. Exact exponents {\it can}, therefore, be found for that problem. But there's no such luck here.

 We`ll now show that any  fixed point of the type (ii)  is unstable under renormalization.

 We start by noting that, in case ii, since $\lambda_1$ is zero, $D_y$ cannot renormalize, since the only nonlinearity is a total $x$-derivative, which cannot generate a term like the $D_y$ term in \rf{angnl2}, which is a total $x$ derivative. Hence, the graphical correction $f_{D_y}$ in the recursion relation \rf{dyrr} for $D_y$ must vanish at the fixed point (ii). Therefore, to keep $D_y$ fixed, \rf{dyrr} implies
\beq
z=2 \,.
\label{dyfix3}
\eeq
Furthermore, the exact recursion relation for the noise strength $\Delta_\theta$ implies that at a fixed point, we must have 
\beq
z-2\chi-1-\zeta=0 \,.
\label{noisefixii}
\eeq
Using our earlier result \rf{dyfix3} that $z=2$ in this condition \rf{noisefixii}, and solving for the roughness exponent $\chi$ gives
\beq
\chi={1-\zeta\over2} \,.
\label{chiii}
\eeq
Since, as we argued earlier, and as vast numerical evidence suggests, $\chi<0$, \rf{chiii} implies that  $\zeta>1$ for a fixed point of type ii. 

Since $\lambda_2$ goes to a non-zero value at this fixed point, we can say, from its recursion relation equation \rf{lambda2rr}, that the fixed point value $f^*_\lambda$ of $f_\lambda$ at such a fixed point must obey
\beq 
2\chi+z-\zeta+f_\lambda=0 \,,
\label{flcond}
\eeq
which, taken together with our result $z=2$ at fixed points of type ii,  implies 
\beq
3-2\zeta=-f^*_\lambda \,.
\label{fl2}
\eeq 

Now, using our results \rf{dyfix3}, \rf{chiii}, and \rf{fl2}   in the recursion relation \rf{lambda1rr} for $\lambda_1$ gives
\beq
\frac{d\lambda_1}{d\ell}=[\chi+z-1+f^*_\lambda]\lambda_1=\left[{3(\zeta-1)\over2}\right]\lambda_1 \,.
\label{l1rrii}
\eeq
Since, as we established earlier,  the negativity of 
$\chi$ requires $\zeta>1$, $3(\zeta-1)/2$ is positive, and therefore, any fixed point of type ii is unstable to any non-zero $\lambda_1$.

Thus, we are left with possibility (iii); that is, fixed points at which both $\lambda_1$ and $\lambda_2$ are non-zero. From equations \rf{lambda1rr} and \rf{lambda2rr}, we see that, at such a  fixed point, we must have
\beq
\chi+z-1+f_\lambda(\lambda_1^*, \lambda_2^*, D_x^*, D_y^*, \Delta_\theta^*)=0 \,,
\label{lambda1fix}
\eeq
\beqn
2\chi+z-\zeta+f_\lambda(\lambda_1^*, \lambda_2^*, D_x^*, D_y^*, \Delta_\theta^*)=0
\label{lambda2fix}
\eeqn

Subtracting \rf{lambda1fix} from \rf{lambda2fix} gives us another scaling relation between the exponents $\chi$ and $\zeta$:
\beq
\chi-\zeta+1=0 \,.
\label{2ndscale}
\eeq

This, taken together with the relation \rf{delthetafix} (which we repeat here for the reader's convenience):
\beq
z-\zeta-2\chi-1=0
\label{1stscale2}
\eeq
gives us two exact scaling laws relating the three universal exponents $\chi$, $\zeta$, and $z$.  These {\emph{quantitative}} predictions could be tested against experiments.

{A further, more qualitative, experimental prediction concerns the form of the static structure factor. Using the standard trajectory integral matching technique \cite{trajec, John_bk}, we can write the static structure factor of angular fluctuations in a form that looks exactly like the one from the linear theory, but with wavevector-dependent quantities $D_x({\bf q})$ and $D_y({\bf q})$ (but a noise strength $\Delta_\theta$ that is wavevector-\emph{in}dependent at small wavenumbers, since it is not renormalised):
\begin{equation}
    {C^{\theta}_{\bf q}}=\frac{\Delta_\theta}{D_x({\bf q})q_x^2+D_y({\bf q})q_y^2}\,.
\end{equation}
Here, the result of integrating out the small wavelength modes is contained in the ``renormalised'' $D_x(\bq)$ and $D_y(\bq)$. We {cannot} calculate these renormalised quantities, but we expect \emph{both} of them to diverge at small wavenumbers (since nothing protects either of them from renormalisation and, as argued earlier, both of them should increase under renormalisation). This implies that $\langle|\theta({\bf q},t)|^2\rangle$ should diverge at small wave numbers with a power that is smaller than $2$ for all wavevector directions.}  This is, of course, consistent with our earlier result that the roughness exponent $\chi<0$.

Note also that this divergence is much weaker, for $\bq$ along the direction $x$ of mean flock motion, than the $1/q^2$ divergence in that direction found by \cite{TT98, Toner_rean}, as pointed out in the introduction.

If we were able to obtain an independent third exact scaling relation, we could go further and 
predict the values of all three of these exponents. This can be done, for example, for  3{D} easy plane Malthusian flocks \cite{toner2024birth}, for which the irrelevance
of the $\lambda_2$ vertex implies that $D_x$ gets no graphical corrections, which provides one with a third exact scaling relation.

Unfortunately, we have been unable to obtain such a third exact scaling relation. Solon and Chate \cite{Solon_Chate} claim that such a relation follows from a ``pseudo-Galilean invariance'' of the equation of motion \eqref{thetaeqmal}. However, the equation actually does not have the symmetry they claim, as we will now show.

Their argument begins by writing the equation of motion \rf{angnl2} before renormalization in the form

\begin{equation}
\label{Gal1}
\partial_t\theta=-\lambda\left(\theta\partial_y\theta-\frac{1}{2}\theta^2{{\partial_{x}}}\theta\right)+D_x{{\partial^2_{x}}}\theta+D_y\partial_y^2\theta+\xi_\theta \,.
\end{equation}

They then assert that this equation has {an invariance under a transformation in which a constant shift of $\theta$ is compensated by a moving shift of the spacetime coordinates.}
That is, if $\theta(x,y,t)$ is a solution of \eqref{Gal1}, there is a family of solutions
$\theta'(x',y',t') {{=}} \theta(x+x_c,y+y_c,t+t_c)+\theta_0$, {for non-constant $x_c,y_c,t_c$}. Replacing $\theta$ by $\theta+\theta_0$ in \eqref{Gal1}, we get
\begin{eqnarray}
\label{Gal2}
\partial_t\theta&=&-\lambda\left[(\theta+\theta_0)\partial_y\theta-\frac{1}{2}(\theta+\theta_0)^2\partial_x\theta\right]\nonumber\\
&&+D_x\partial_x^2\theta+D_y\partial_y^2\theta+\xi_\theta.
\end{eqnarray}
From the form of \eqref{Gal2}, at first sight, it seems that taking
\begin{equation}
\label{transform}
y_c=-\lambda\theta_0 t, \, x_c=\lambda\left[\theta_0\int^t_0\theta d\bar{t}+\frac{\theta_0^2 t}{2}\right],\, t_c=0
\end{equation}
 then recovers our original equation, which suggests 
that the equation of motion \rf{angnl2} does have a pseudo-Galilean {invariance. Specifically,  t}he time-derivative becomes
\begin{equation}
\partial_t\theta\equiv\partial_{t'}\theta-\lambda\left[\theta_0\partial_{y'}-\left(\theta_0\theta+\frac{\theta_0^2}{2}\right)\partial_{x'}\right]\theta,
\end{equation}
apparently cancelling the $\theta_0$-dependent parts on the R.H.S. of \eqref{Gal2}. 
 
However, because their transformation of the coordinates involves the field $\theta$, which is a function of space and time, and not simply the parameter $\theta_0$ and time, $\partial_{x'}\theta\neq\partial_{x}\theta$,  and $\partial_{y'}{{\theta}}\neq\partial_y\theta$, in contrast to the usual Galilean transformations \cite{Toner_mal}. 
 Instead,
\begin{equation}
\partial_{x}\theta=\left(1+\lambda\theta_0\partial_{x}\int_0^t\theta d\bar{t}\right)\partial_{x'}\theta
\end{equation}
and
\begin{equation}
    \partial_{y}\theta=\partial_{y'}\theta+\left(\lambda\theta_0\partial_y\int_0^t\theta d\bar{t}\right)\partial_{x'}\theta.
\end{equation}
This implies that the terms on the R.H.S. of \eqref{Gal2} do not remain unchanged upon the transformation of coordinates to $(x',y',t')$; instead they transform in a complicated, field-dependent way. 

As a result,  it is incorrect to argue, as \cite{Solon_Chate} do, that the equation possesses {the claimed} pseudo-Galilean invariance which protects $\lambda_{1,2}$ from graphical renormalization. In fact, we expect that $\lambda_{1,2}$ {\it will} get graphical renormalizations (i.e., that $f_\lambda$ in equations \rf{lambda1rr} and \rf{lambda2rr} is non-zero. As a result, we cannot obtain a third exact scaling relation, and, hence, have no simple way to determine the exact values of the exponents $\chi$, $\zeta$, and $z$ for two-dimensional Malthusian flocks.

Nor have we been able to find any alternative approach to determining these exponents, even approximately. The obvious approach of actually calculating the graphical corrections can, unfortunately, only be done perturbatively in the nonlinearities. This has been done for our original model equations \rf{vEOM} and \rf{rhoeq} with the birth and death term $h(\rho)\ne0$ in a $d=4-\epsilon$-expansion by \cite{Mal_RG1, Mal_RG2}, who obtained exponents that are quite trustworthy in three dimensions{; it was not necessary to take the $\lambda_2$ nonlinearity into account for this calculation since it is \emph{irrelevant} near $d=4$ i.e., the fixed point for dimensions near $4$ is of type (i)\footnote{{Using the dimension-dependent versions of the linear exponents $\chi_{\text{lin}}$, $\zeta_{\text{lin}}$, $z_{\text{lin}}$  in Eq. \eqref{lambda2rr} with $f_\lambda=0$, we find that $\lambda_2$ becomes relevant for $d<3$. If we instead use the $\mathcal{O}(\epsilon)$ result of the $d=4-\epsilon$ calculation of \cite{Mal_RG1, Mal_RG2} even for $\epsilon>1$, we find that it becomes relevant for $d<25/9$.}}}. However, the perturbation theory, as usual in $\epsilon$ expansions, gets less reliable as one goes to lower spatial dimensions, and certainly is completely untrustworthy by the time one reaches $d=2$. So we unfortunately can say nothing quantitative about the values of the exponents for 2d Malthusian flocks, other than that they obey the two exact scaling relations \rf{2ndscale}and \rf{1stscale2}.

{\subsection{Justification for ignoring density fluctuations in the Malthusian flock}}

{We now show that density fluctuations in Malthusian flocks can be ignored, as we have done in this section. Retaining only the most relevant terms, the dynamical equation for density fluctuations $\delta\rho$ about its steady state value $\rho_0$ is given by
\begin{equation}
\partial_t\delta\rho=-\beta(\rho_0)\partial_y\theta-h'(\rho_0)\delta\rho+\xi_n\,.
\end{equation}
From this, we can solve for $\delta\rho$:
\begin{equation}
\label{delrhoapp}
\delta\rho\approx\frac{\xi_n-\beta(\rho_0)\partial_y\theta}{h'(\rho_0)}\,,
\end{equation}
We now examine how density fluctuations affect the $\theta$ equation. To do this, we have to obtain the terms involving the density field in that equation. We do this in detail in Sec. \ref{eomimmort}. Anticipating that discussion, we see that we have (i) a linear term $\propto c_s\partial_y\delta\rho$, (ii) nonlinear terms of the form $\partial_y(\delta\rho)^2$ and (iii) nonlinear terms that involve some power(s) of $\theta$, one gradient and $\delta\rho$. Representative examples are $\theta\partial_x\delta\rho$, $\delta\rho\partial_x\theta$, $\theta^2\partial_y\delta\rho$ etc. Using Eq. \eqref{delrhoapp}, we find that the only effect of the linear term $c_s\partial_y\delta\rho$ is to change the value of $D_y$  by a {\it finite} amount. However, since $D_y$ was a phenomenological coefficient, to begin with, this doesn't change our previous conclusions. }

{Ignoring graphical corrections, the coefficients of nonlinearities of type (ii) above (i.e., those $\propto\partial_y(\delta\rho)^2\sim\partial_y(\partial_y\theta)^2$) scale as $b^{z-3+\chi}$, rendering them irrelevant. }

{Similarly, using \eqref{delrhoapp} to replace $\delta\rho$ in nonlinearities of type (iii), we find that they yield terms nonlinear in $\theta$ that involve \emph{two} derivatives; we have ignored such terms from the start. Indeed, the coefficients of these nonlinearities scale either as $b^{z-1-\zeta+\chi}$ or as $b^{z-2+2\chi}$ and are less relevant than the families of nonlinearities in \eqref{thetaeqmalexp}. }\\

\section{Immortal flocks}{\label{Immortal}}

\subsection{Equation of motion}{\label{eomimmort}}

We turn now to the long-time, large-scale dynamics of the ordered phase of an {\it immortal} flock, which we remind the reader is a flock in which the number of flockers is conserved.  In the absence of noise, this ordered state will
simply be any one of the infinity of uniform steady states \eqref{ss} of the equations of motion \rf{vEOM} and \rf{rhoeq}
\beq
\bp(\br,t)=p_0 \hn \sep \rho(\br,t)=\rho_0={\rm constant}\,,
\label{ss2}
\eeq
where $p_0$ is the value of the magnitude of the polarization at which $U({\rho_0}, |\bp|)$ vanishes.

Precisely as we did for Malthusian flocks, we now study fluctuations about this uniform state by writing the polarization as
\beqn
{\bf p(\br,t)}=(p_0+\delta p(\br,t))(\cos\theta(\br,t),\sin\theta(\br,t)) \,,
\label{pexp}
\eeqn
\noindent and expand the density around its mean value $\rho_0$:
\beq
\rho(\br,t)=\rho_0+\delta\rho(\br,t) \,.
\label{rhoexp}
\eeq

{Paralleling our discussion for} {Malthusian flocks, we expand \eqref{vEOM} to first order in fluctuations $\delta p(\br,t)$ in the magnitude of the polarization $\bp$, and $\delta\rho$ to obtain}
\bew
\beq
(\cos\theta, \sin\theta)\pp_t\delta p+{p_0}(-\sin\theta, \cos\theta)\pp_t\theta=-(\alpha\delta p+\bar{\alpha}\delta\rho)(\cos\theta, \sin\theta)+... \,,
\label{pthetarhoeom}
\eeq
\ew
{where we've defined $\alpha\equiv -p_0\left({\pp U(\rho, |\bp|)\over\pp |\bp|}\right)_{\rho_0, p_0}$ and $\bar{\alpha}\equiv -p_0\left({\pp U(\rho, |\bp|)\over\pp \delta\rho}\right)_{\rho_0, p_0}$ and the ellipsis denotes terms involving spatial derivatives of $\theta$,
$\delta p$, and $\delta\rho$ as well as products of $\delta p$, $\delta\rho$ and spatial derivatives of $\theta$. Projecting this along {$\hn$}, here too we find that $\delta p$ fluctuations are ``fast'', and can therefore be eliminated by solving its equation of motion in the long-time limit, and inserting the result into the equations of motion for {the fields $\theta(\br,t)$ and $\dr(\br,t)$. }}

The result {of this calculation} is a closed set of equations of motion for those two fields, whose form is required by the symmetry of rotation invariance, and the conservation of particle number. {We directly write these down without going through the algebra described in Sec. \ref{eommalt} to obtain them from \eqref{vEOM} and \eqref{rhoeq}.} These {equations} read:

\begin{multline}
	\label{rhoeq2}
	\partial_t\rho=-\nabla\cdot[(\beta_0+w_1\delta\rho+w_2\delta\rho^2)(\cos\theta,\sin\theta)]\\+D_{\theta\rho}\partial_x\partial_y\theta+D_{ x\rho}\partial_x^2\delta\rho+D_{y\rho}\partial_y^2\delta\rho+\nabla\cdot\boldsymbol{\xi}_\rho,
\end{multline}
\begin{multline}
	\label{thetaeq}
	\partial_t\theta=(\lambda+g_1\delta\rho)\nabla\cdot(-\sin\theta,\cos\theta)+D_x\partial_x^2\theta+D_y\partial_y^2\theta\\-(-\sin\theta,\cos\theta)\cdot\nabla({c_s}\delta\rho+g_3\delta\rho^2)+D_{{ \rho\theta}}\partial_x\partial_y\delta\rho+\xi_\theta,
\end{multline}
where the Gaussian, zero-mean white noises have correlations
\beqn
\langle{\xi}_\theta({\bf x}, t){\xi}_\theta({\bf x}', t')\rangle&=&2{\Delta}_\theta\delta({\bf x}-{\bf x}')\delta(t-t') \,, \\
\langle{\xi}_{\rho i}({\bf x}, t){\xi}_{\rho j}({\bf x}', t')\rangle&=&2{\Delta}_\rho\delta_{ij}\delta({\bf x}-{\bf x}')\delta(t-t') \,.
\label{noisecorrimm}
\eeqn
In writing these equations, we have stopped the expansion in powers of $\delta\rho$ at $O(\delta\rho^2)$, in anticipation of finding a negative value of the density roughness exponent $\chi_\rho$, which will make higher powers of $\delta\rho$ less relevant. 

{
We can think of $\beta_0$, $w_1$, and $w_2$ as the expansion coefficients of $\beta$ in equation \rf{rhoeq}.}

{These equations are invariant under the symmetry transformation in Eq. \eqref{rotinv}, by construction. Note that the dynamical equation for the Nambu-Goldstone mode in \cite{boost_agnostic} is not symmetric under \eqref{rotinv} and, therefore, it is incorrect. Further, arguments of Ref. \cite{boost_agnostic} for asserting that various nonlinear terms are not renormalised are also incorrect. In fact, as we will discuss below, unlike in Malthusian flocks, we cannot find \emph{any} hyperscaling relation in immortal flocks.} 

Note {that} the time derivative of $\theta$ as given by \rf{thetaeq} is {\it not} a total divergence. 
{As we noted in section \rf{malt}, and contrary to the claim in reference \cite{Solon_Chate}, there is no reason why it, or in general the rate of change of any Nambu-Goldstone mode, should be one.} 
Hence, \cite{Solon_Chate}'s starting equation of motion for the immortal case, and, therefore, the exponents they derive from it, are incorrect.
{Furthermore, {comparison with \eqref{thetaeq} suggests that}
their claimed numerical generation of a term $\propto\theta$, in the $\theta$ equation, by simulating Eq. (22), (23) of their SI (see Fig. S4 of \cite{Solon_Chate}) {is a result of inadvertently breaking \emph{rotation} symmetry. Our analysis makes it clear that for Chat\'{e} and Solon's \cite{Solon_Chate} Eq. 18 to be rotation invariant, their $h_{x,2}$ has to be equal to $\lambda_1$ (or equivalently, $h_{x,2} = -\tilde{\sigma}$ in (22), (23) in their SI); see Eq. \eqref{rotinv}, and Eqs. \eqref{rhoeq2} and \eqref{thetaeq} of this article.  Indeed it is only for that case that their Fig. S4 shows a massless correlator. Fig. S4 shows that other choices of $h_{x,2}$ give a mass to their $\phi$ field, unsurprisingly, since they break rotation invariance regardless of the value of $h_{x1}$.
}As all nonlinear terms in Eqs. \eqref{rhoeq2} and \eqref{thetaeq} are invariant under the transformation in Eq. \eqref{rotinv} and a term $\propto\theta$ is not, no such term can be generated.}

\subsection{Linear theory}{\label{immortlin}}

The linear theory of the immortal flock is discussed in detail in \cite{Toner_rean}. The key results of this analysis are as follows: 
{the eigenfrequencies $\omega_\pm(\bq)$ of modes with wavevector $\bq= q(\cos\phi,\sin\phi)$,} where $\phi$ is the angle between the wavevector ${\bf q}$ and the ordering direction $\hat{x}$, {are} given by
\beq
\omega_\pm=c_\pm(\phi) q-iD_\pm(\phi) q^2 \,
\label{dispimmort}
\eeq
with the direction $\phi$-dependent sound speeds $c_\pm(\phi)$ and dampings $D_\pm(\phi)$ given by
\begin{equation}
	c_\pm(\phi)=\left(\frac{w_1+\lambda}{2}\right)\cos\phi\pm\sqrt{\frac{(w_1-\lambda)^2}{4}\cos^2\phi+{c_s}\beta\sin^2\phi} \,,
	\label{soundspeed}
\end{equation}
\bew
\begin{multline}
	D_\pm(\phi)=\left(\frac{D_x+D_{x\rho}}{2}\right)\cos^2\phi+\left(\frac{D_y+D_{y\rho}}{2}\right)\sin^2\phi
	\\
\pm{1\over4}\left(\frac{\cos\phi}{c_\pm(\phi)+(w_1+\lambda)\cos\phi}\right)
\bigg[\bigg\{(D_{y\rho}-D_y)\sin^2\phi+(D_{x\rho}-D_x)\cos^2\phi\bigg\}(w_1-\lambda)+2(\beta D_{{\rho\theta}}+{c_s} D_{{\theta\rho}})\sin^2\phi\bigg]\,.
\label{damping}
\end{multline}
\ew

In contrast to the Malthusian case, the wavevector dependence \rf{dispimmort} of the eigenfrequencies appears to leave some ambiguity as to the value of the dynamical exponent $z$: the linear dependence of the real ($c_\pm$) part suggests $z=1$, while the quadratic form of the imaginary ($D_\pm$) part implies $z=2$. However, since it is the imaginary part (i.e., the damping) that controls the scale of the angular fluctuations (as we'll see explicitly in a moment), the choice $z=2$ is the correct one for the linear theory.

To see this, we begin with the {spatio-temporal Fourier transform of the correlation functions $C^\rho({\bf r},t)\equiv\langle\rho({\bf 0},0)\rho({\bf r},t)\rangle$ and $C^\theta({\bf r},t)\equiv\langle\theta({\bf 0},0)\theta({\bf r},t)\rangle$} obtained from the linear theory: 
\begin{equation}
    {C^\rho_{\bf q,\omega}}
    =\frac{\beta^2q^2\sin^2\phi\Delta_\theta}{[(\omega-c_-q)^2+D_-^2q^4][(\omega-c_+q)^2+D_+^2q^4]}
\end{equation}
\begin{equation}
    {C^\theta_{\bf q,\omega}}
    =\frac{(\omega-w_1q\cos\phi)^2\Delta_\theta}{[(\omega-c_-q)^2+D_-^2q^4][(\omega-c_+q)^2+D_+^2q^4]}
\end{equation}
Integrating these overall frequency gives the spatially Fourier-transformed equal-time correlation functions:
\begin{equation}
    {C^\rho_{\bf q}}
    =\frac{\beta^2(D_++D_-)\sin^2\phi\Delta_\theta}{2q^2(c_+-c_-)^2D_+D_-}
\end{equation}
\begin{equation}
    {C^\theta_{\bf q}}
    =\frac{[D_-(c_+-w_1\cos\phi)^{2}+D_+(c_--w_1\cos\phi)^{2}]\Delta_\theta}{2q^2(c_+-c_-)^2D_+D_-}.
\end{equation}
  From these, we can see that it is the imaginary (i.e., the damping) parts $D_\pm$ of the eigenfrequencies that control the scale of the fluctuations. To make this clear, note that the equal-time correlation functions only depend on ratios of the speeds $\beta$, {$c_s$}, $w_1$, and $\lambda$, not on the absolute values of those speeds themselves.
Thus, the scale of the fluctuations remains fixed if we keep the diffusion constants $D_x$, $D_y$, $D_{x\rho}$, $D_{y\rho}$, $D_{\rho\theta}$, and $D_{\theta\rho}$ fixed. This implies that the linear theory will have the dynamical exponent $z=2$.

Since the spatially Fourier transformed density $\rho(\bq,t)$ and angle $\theta(\bq,t)$ mean squared fluctuations both scale as $1\over q^2$, the mean squared real space fluctuations of both, in the linear theory, will scale in two dimensions like  $\ln(L)$, where $L$ is the system size, just as the angle fluctuations do in the linear theory of Malthusian flocks studied in section \rf{malt}. Hence, the ``roughness'' exponents $\chi$ and $\chi_\rho$ for the fields $\theta$ and $\delta\rho$, respectively, are $\chi=\chi_\rho=0$ in the linear theory.

The scaling of angular fluctuations of both immortal and Malthusian flocks within a linear theory are equivalent to those of equilibrium polar or nematic liquid crystals. Unlike in equilibrium systems (away from a critical point), active currents make the density fluctuations as large as the Nambu-Goldstone fluctuations \cite{Aditi_2, Toner_shocking}. These linear results are well-established. It is also well-known that they are modified by nonlinearities. We will re-examine the effect of nonlinearities on the scaling of hydrodynamic correlators via the standard renormalisation group procedure.

\subsection{Dynamical Renormalization group for immortal flocks}{\label{immortdrg}}

The dynamical renormalization group analysis of the equations of motion \rf{rhoeq2} and \rf{thetaeq} proceeds exactly as in the Malthusian case discussed in section \ref{malt}. The only difference is that now we must 
{consider} the density field as well. Our rescalings are:
\beq
y\to b y \sep t\to b^z t \sep x\to b^\zeta x \sep  \theta\to b^\chi\theta \sep \rho\to b^{\chi_\rho}\rho \,.
\label{rescale2}
\eeq
As with the Malthusian problem, we once again begin by
 expanding the equations of motion \rf{rhoeq2} and \rf{thetaeq} for $\delta\rho$ and  $\theta$ in powers of $\theta$. This is easily seen to lead to 
 \bew
\beqn
\partial_t\rho&=&\sum_{n=1}^\infty k_n\pp_x(\theta^{2n})+\sum_{n=0}^\infty \bigg(v_n\pp_y(\theta^{2n+1})+d_n\pp_x(\theta^{2n}\dr)+f_n\pp_y(\theta^{2n+1}\dr)+{s_n}\pp_x(\theta^{2n}\dr^2)+h_n\pp_y(\theta^{2n+1}\dr^2)\bigg) \nn\\
	&&\,\,\,\, +D_{\theta\rho}\partial_x\partial_y\theta+D_{x\rho}\partial_x^2\delta\rho+D_{y\rho}\partial_y^2\delta\rho+\nabla\cdot\boldsymbol{\xi}_\rho
		\label{rhoeqimmexp}
		\\
		\partial_t\theta&=&\sum_{n=0}^\infty \bigg(a_n\theta^{2n}\pp_x\theta+c_n\theta^{2n+1}\pp_y\theta+\Upsilon_n\dr\,\theta^{2n}\pp_x\theta+\varXi_n\dr\,\theta^{2n+1}\pp_y\theta+\gamma_n\theta^{2n+1}\pp_x\dr+e_n\theta^{2n}\pp_y\dr
				\nn\\
	&&\,\,\,\, +\mu_n\theta^{2n+1}\pp_x(\dr^2)+\sigma_n\theta^{2n}\pp_y(\dr^2)\bigg)
	+D_x\partial_x^2\theta+D_y\partial_y^2\theta
	+D_{{\rho\theta}}\partial_x\partial_y\delta\rho+\xi_\theta \,,
	\label{thetaeqimmexp}
\eeqn
\ew
where the expansion coefficients before we begin renormalizing (i.e., at DRG ``time''
$\ell=0$,  are given by 
\begin{subequations}
\begin{equation}
k_n(\ell=0)=-{(-1)^n\beta\over(2n)!} \,,
\end{equation}
\begin{equation}
v_n(\ell=0)=-{(-1)^n\beta\over(2n+1)!} \,,
\end{equation}
\begin{equation}
d_n(\ell=0)=-{(-1)^nw_1\over(2n)!}\,,
\end{equation}
\begin{equation}
f_n(\ell=0)=-{(-1)^nw_1\over(2n+1)!} \,,
\end{equation}
\begin{equation}
{s_n}(\ell=0)=-{(-1)^nw_2\over(2n)!} \,,
\end{equation}
\begin{equation}
h_n(\ell=0)={-}{(-1)^nw_2\over(2n+1)!} \,,
\end{equation}
\begin{equation}
c_n(\ell=0)=-{(-1)^n\lambda\over(2n+1)!} \,,
\end{equation}
\begin{equation}
a_n(\ell=0)=-{(-1)^n\lambda\over(2n)!} \,,
\end{equation}
\begin{equation}
\Upsilon_n(\ell=0)=-{(-1)^ng_1\over (2n)!}\,,
\end{equation}
\begin{equation}
\varXi_n(\ell=0)= -{(-1)^ng_1\over(2n+1)!} \,,
\end{equation}
\begin{equation}
\gamma_n(\ell=0)={(-1)^n{c_s}\over(2n+1)!} \,,
\end{equation}
\begin{equation}
e_n(\ell=0)=-{(-1)^n{c_s}\over(2n)!}\,,
\end{equation}
\begin{equation}
\mu_n(\ell=0)={(-1)^ng_3\over(2n+1)!} \,,
\end{equation}
\begin{equation}
\sigma_n(\ell=0)=-{(-1)^ng_3\over(2n)!}\,.
\end{equation}
\label{bareexp2}
\end{subequations}
Note that the parameters $v_0$, $c_0$, and $e_0$ in \rf{thetaeqimmexp} and \rf{rhoeqimmexp} are precisely the parameters $-\beta$, $-\lambda$, and $-c_s$ appearing in the linear theory of section \rf{immortlin}.

As with the Malthusian problem, these expressions will not continue to hold as we iterate the DRG. However, once again  it is only the rescaling step of the DRG that changes the ratios 
  $k_n/k_{n^\prime}$, $v_n/v_{n^\prime}$, and, most importantly,
$v_n/k_{n^\prime}$, because, before rescaling, these terms must resum to the
$\nabla\cdot[(\beta(\cos\theta,\sin\theta)]$ term in \rf{rhoeqimmexp}. Likewise, if one replaces $k_n$
and $v_n$ with the pairs $(d_n, f_n)$, $({s_n}, h_n)$, {$(a_n, c_n)$}, $(\Upsilon_n,
\varXi_n)$, {$(\gamma_n, e_n)$}, and $(\mu_n, \sigma_n)$ in the previous sentence, it
remains true; that is, those ratios remain fixed before rescaling as well.

The recursion relations between the parameters of \rf{rhoeqimmexp} and \rf{thetaeqimmexp} on successive steps of the RG are
    \begin{subequations}
\begin{equation} {k}_n^\prime= b^{2n\chi-\chi_\rho+z-\zeta} ({k}_n + {\rm graphs}) \,, \label{anresim}
\end{equation}
\begin{equation}
 v_n^\prime= b^{(2n+1)\chi-\chi_\rho+z-{-1}} (v_n + {\rm graphs}) \,, \label{cnresim}
 \end{equation}
 \begin{equation}
  d_n^\prime= b^{2n\chi+z-\zeta} (d_n + {\rm graphs}) \,, \label{dnres}
  \end{equation}
  \begin{equation}
 f_n^\prime= b^{(2n+1)\chi+z-1} (f_n + {\rm graphs}) \,, \label{fnres}
 \end{equation}
 \begin{equation}
{s_n}^\prime= b^{2n\chi+\chi_\rho+z-\zeta} ({s_n} + {\rm graphs}) \,, \label{gnres}
 \end{equation}
 \begin{equation}
 h_n^\prime= b^{(2n+1)\chi+\chi_\rho+z-1} (h_n + {\rm graphs}) \,, \label{hnres}
 \end{equation}
 \begin{equation}
{a}_n^\prime= b^{2n\chi+z-\zeta} ({a}_n + {\rm graphs}) \,, \label{alphanres}
 \end{equation}
 \begin{equation}
{c}_n^\prime= b^{(2n+1)\chi+z-1} ({c}_n + {\rm graphs}) \,, \label{betanres}
 \end{equation}
 \begin{equation}
\Upsilon_n^\prime= b^{2n\chi+\chi_\rho+z-\zeta} (\Upsilon_n + {\rm graphs}) \,, \label{upsnres}
\end{equation}
\begin{equation}
 \varXi_n^\prime= b^{(2n+1)\chi+\chi_\rho+z-1} (\varXi_n + {\rm graphs}) \,, \label{xinres}
 \end{equation}
 \begin{equation}
  \gamma_n^\prime= b^{2n\chi+\chi_\rho+z-\zeta} (\gamma_n + {\rm graphs}) \,, \label{gammanres}
  \end{equation}
  \begin{equation}
 {e}_n^\prime= b^{(2n-1)\chi+\chi_\rho+z-1} ({e}_n + {\rm graphs}) \,, \label{nunres}
 \end{equation}
 \begin{equation}
 \mu_n^\prime= b^{2n\chi+2\chi_\rho+z-\zeta} (\mu_n + {\rm graphs}) \,, \label{munres}
 \end{equation}
 \begin{equation}
 \sigma_n^\prime= b^{(2n-1)\chi+2\chi_\rho+z-1} (\sigma_n + {\rm graphs}) \,, \label{sigmares}
 \end{equation}
 \begin{equation}
 D_{\theta\rho}^\prime=b^{z+\chi-\chi_\rho-1-\zeta}(D_{\theta\rho}+ {\rm graphs}) \,, \label{dthetarhores}
 \end{equation}
 \begin{equation}
 D_{x\rho}^\prime=b^{z-2\zeta}(D_{x\rho}+ {\rm graphs}) \,, \label{dxrhoresim}
 \end{equation}
 \begin{equation}
 D_{y\rho}^\prime=b^{z-2}(D_{y\rho}+ {\rm graphs}) \,, \label{dyrhoresim}
 \end{equation}
 \begin{equation}
D_x^\prime=b^{z-2\zeta}(D_x+ {\rm graphs}) \,, \label{dxresim}
\end{equation}
\begin{equation}
 D_y^\prime=b^{z-2}(D_y+ {\rm graphs}) \,, \label{dyresim}
 \end{equation}
 \begin{equation}
  D_{\rho\theta}^\prime=b^{z+\chi_\rho-\chi-\zeta-1}(D_{\rho\theta}+ {\rm graphs}) \,, \label{drhothetares}
  \end{equation}
  \begin{equation}
   \Delta_\rho^\prime=b^{z-2\chi_\rho-\zeta-1}(\Delta_\rho+ {\rm graphs}) \,,\label{delrhoresim}
   \end{equation}
   \begin{equation}
 \Delta_\theta^\prime=b^{z-2\chi-\zeta-1}(\Delta_\theta+ {\rm graphs}) \,,\label{delthetaresim}
 \end{equation}
 \end{subequations}
where the primes denote parameters after the RG step in question, and ``graphs'' denote the renormalizations arising from averaging over the short-wavelength degrees of freedom. 

Again as we did for the Malthusian problem, here too we can use these recursion relations, even without evaluating the graphical corrections,  to determine which of the nonlinearities {$k_n$}, $v_n$,  $d_n$, $f_n$, {$s_n$}, $h_n$,  {$a_n$, $c_n$},
$\Upsilon_n$, $\varXi_n$,  $\gamma_n$, {$e_n$}, $\mu_n$, and $\sigma_n$ are ``relevant'' in the RG sense.  We do this by once again noting that in the absence of those nonlinearities, the graphical corrections must vanish.

Hence, if we choose the rescaling exponents $z$, $\zeta$, and $\chi$ in \rf{dthetarhores}-\rf{delthetaresim} to keep the
linear damping  terms  $D_x$, $D_y$, $D_{\theta\rho}$, $D_{x\rho}$, $D_{y\rho}$, and
$D_{\rho\theta}$,  and the noise strengths $\Delta_\theta$ and $\Delta_\rho$ fixed (so that
the scale of the fluctuations remains the same upon renormalization), then we can assess
the relevance or irrelevance of the nonlinearities {$k_n$} and $v_n$ when those nonlinearities are initially small (so that the graphical corrections can be ignored in the early
RG steps)  by asking simply whether or not they grow upon renormalization if they start out being small.

Ignoring the graphical corrections, we see that equations \rf{dxresim},\rf{dyresim}, \rf{delrhoresim}, and \rf{delthetaresim} imply that,  to keep the linear damping  terms  $D_x$, $D_y$,
and the noise strengths
$\Delta_\theta$ and $\Delta_\rho$ fixed, we must choose the rescaling exponents $z$, $
\zeta$, $\chi_\rho$, and $\chi$ so that the following four conditions are satisfied:
\beqn
&z&-2=0 \sep z-2\zeta=0 \sep z-\zeta-2\chi-1=0
\nn\\
&z&-\zeta-2\chi_\rho-1=0
\,,
\label{lincond2}
\eeqn
whose solutions are trivially found to
\beq
z=2 \sep \zeta=1 \sep \chi=\chi_\rho=0 \,,
\label{zlinrg2}
\eeq
which the alert reader will recognize as the exponents we found earlier by inspecting the correlation function predicted by the linear theory.

It is also straightforward to see that these exponents also keep the remaining four linear damping parameters  
$D_{\theta\rho}$, $D_{x\rho}$, $D_{y\rho}$, and $D_{\rho\theta}$ fixed.

Using these exponents in 
 \rf{anresim}-\rf{sigmares}, we find that every single nonlinear term {$\nu_n$}, where {$\nu_n$} can stand for any one of the parameters {$k_n$} through
$\sigma_n$ in equations \rf{anresim}-\rf{sigmares}
obeys an identical recursion relation, which reads
\beqn
{\nu_n}^\prime= b({\nu_n} + {\rm graphs}) \,, \label{wnlinexp}
\eeqn
We see from this immediately that {\it all} of the double septuplets of infinities of
nonlinearities {$k_n$} through $\sigma_n$  are {\it relevant} (that is, grow upon renormalization) if they are initially small (so that the ``graphs'' can be neglected). This proves that the linear results are {\it never} valid in a two-dimensional immortal flock, except in the very special case in which every single one of the aforementioned double septuplets of infinities of
nonlinearities $k_n$ through $\sigma_n$ vanish, which will never occur in practice.

The long wavelength physics will therefore always be changed by the nonlinearities and must be controlled by an RG fixed point at which at least one of the nonlinearities  {$\nu_n$}
is non-zero.

We'll now argue that only a finite (although, alas, distressingly large) subset of the infinity of nonlinearities in \rf{rhoeqimmexp} and \rf{thetaeqimmexp} actually need to be considered.

The argument for this is basically the same as that used in our discussion in section \rf{malt} of Malthusian flocks and relies on arguing that the roughness exponents $\chi$ and $\chi_\rho$ are both less than zero.

Unfortunately, for reasons that we'll discuss in more detail in a moment, we do not have as compelling an argument for the negativity of $\chi$ and $\chi_\rho$ as we had for the negativity of $\chi$ in Malthusian flocks. However, it remains true that, as we noted in our discussion of Malthusian flocks, that $\chi$ must be negative in order to have a long-range ordered state. And $\chi_\rho$ must be negative, 
as can be seen by the following {\it reductio ad absurdum} argument:
The renormalization group implies that  
\beq
    P(\delta\rho, L)={L^{-\chi_\rho}}f(\delta\rho/L^{\chi_\rho}) \,,
\label{prho}
\eeq
where $\delta \rho$ is the departure of the space averaged value of $\rho$ in a cubic ``sample box'' of side $L$ from its mean value $\rho_0$, and $P(\delta \rho, L)$ is the probability distribution of that mean density departure. Hence, the mean value $\langle\rho\rangle$ of $\rho$ in that sample box is given by
\beq
    \langle\rho\rangle=\rho_0+\int_{-\rho_0}^\infty f(\delta \rho/L^{\chi_\rho}) \,\delta \rho\, d\, \delta \rho \,,
\label{rhobar}
\eeq
where $f(x)$ is a positive definite scaling function. Changing variables of integration to $x=\delta \rho/L^{\chi_\rho}$, and assuming (as we will now prove by contradiction is impossible) that $\chi_\rho>0 $, we obtain, for  $L$ very large,
\beqn
    \langle\rho\rangle&=&\rho_0+L^{\chi_\rho} \int_{-\rho_0/L^{\chi_\rho}}^\infty f(x) x dx \nn\\&\approx& \rho_0+L^{\chi_\rho} \int_0^\infty f(x) x dx \,,
\label{rhobarscale}
\eeqn
where the approximate equality becomes asymptotically exact in the limit $L\to\infty$.

The $ \int_0^\infty f(x) x dx$  in the last expression is positive definite, since $x>0$ throughout the range of integration, and $f(x)$ is positive definite, being a probability distribution.

 Since in the limit of $L\rightarrow \infty$ the left-hand side is equal to $\rho_0$ by definition, the second term on the right-hand side must vanish in the same limit, which is in contradiction with the assumption $\chi_\rho>0$. Hence, $\chi_\rho<0$.




Furthermore, the fact that the graphical corrections to the diffusion constants $D_{x,y,(x\rho), (y\rho), (\theta\rho), (\rho\theta)}$ are probably all positive tends to reduce $\chi$ and
$\chi_\rho$ from zero. Unfortunately, unlike the Malthusian case, here (as we'll show in a moment) the noise strengths $\Delta_\theta$ and $\Delta_\rho$ are also renormalized, and it's not obvious {\it a priori} that the increase in the fluctuations caused by this enhanced noise doesn't overwhelm the suppression of the fluctuations caused by the enhancement of the damping.

However, given the extensive numerical evidence \cite{chate2020dry} that a long-range ordered state {\it does} exist in immortal flocks, it seems safe to conclude that $\chi$ and $\chi_\rho$ {\it are} both negative. Given this, it can
 be immediately seen from
the recursion relations \rf{anresim}-\rf{sigmares} that the most relevant nonlinear terms of the set of {$\nu_n$'}s are those with the smallest value of $n$; that is, ${k}_1$, $v_1$, $d_1$, ${a}_1$, and ${e}_1$,  and the $n=0$ term of all the other parameters {$c_n$}, $f_n$, ... $\gamma_n$, $\sigma_n$. We have not included $v_0=-\beta$, ${a_0=-\lambda}$,  $d_0=-w_1$, and $e_0=-c_s$ in this list, since these are linear terms.

So the minimal theory for two-dimensional immortal flocks does not have an infinite number of potentially relevant nonlinearities. However, its finite number is still quite large (fourteen)\footnote{{In a theory of the coupled density and rotational Nambu-Goldstone mode in arbitrary dimensions, seven of these nonlinearities become relevant below $d=4$ as shown in \cite{Toner_rean}, therefore even a perturbative calculation of the exponents near $d=4$ is prohibitively difficult.}}

{In practice, perturbative calculations of the graphical corrections are reliable if the nonlinearities are small at the fixed point.} As in the Malthusian problem, we only expect this to be true near the critical dimension for this problem, which is well-known \cite{TT95,TT98,Toner_rean, John_bk} to be four. Hence, accurate calculation of the graphical corrections in two dimensions is not only formidable (because of the large number of relevant nonlinearities), but actually impossible{, barring non-perturbative miracles}.

Furthermore, the situation for immortal flocks is actually somewhat worse than that for Malthusian flocks, because in the immortal case, we cannot 
{obtain} {\it any} exact scaling relation between the exponents. This is because, as we've now pointed out many times, not all of the nonlinearities in the equation of motion \rf{thetaeqimmexp} for $\theta$ can be written as total derivatives. This 
{rules out the possibility of using our argument from the Malthusian case} that $\Delta_\theta$ gets no graphical renormalization, which eliminates the corresponding exact exponent relation \rf{delthetafix}. Note that \cite{Solon_Chate} assert that this relation {\it does} hold, because they erroneously require that the time derivative of $\theta$ must be a total divergence, because of their mistaken belief that all Nambu-Goldstone modes must have this property. {In fact, 
the relation} \rf{delthetafix} need not hold, and {almost certainly} does not.

The nonlinearities in the $\delta\rho$ equation of motion \rf{rhoeqimmexp} {\it are}, of course, necessarily total divergences, since, in an immortal flock, {the space integral of} $\delta\rho$ is conserved. 
{However, as the noise in \rf{rhoeqimmexp} is conserving, the fact that the nonlinearities are total divergences does not protect $\Delta_{\rho}$ from renormalization.}

We can't even obtain exact scaling  relations like  \rf{2ndscale} that we derived in the Malthusian
case by using the fact that the graphical corrections to $\lambda_1$ and $\lambda_2$
are equal. The reason we can't use the analogous argument for the seven pairs of nonlinearities here that must, pairwise, have identical graphical corrections, is that we do not know that both members of any given pair are non-zero at the fixed point. For example, if we {\it did} know that, say, {$d_1$} and $f_0$ were both non-zero at the fixed point, then equations \rf{dnres} and \rf{fnres}, taken together with the observation that the graphical corrections to both are equal, would imply
\beq
2\chi+z-\zeta=\chi+z-1 \,,
\label{wrongscale}
\eeq
which would immediately imply $\zeta=1{+\chi}$. Unfortunately, since we do {\it not} know that both {$d_1$} and $f_0$ are non-zero at the fixed point, an alternative possibility is that $\zeta<1{+\chi}$, $f_0$ is irrelevant at the fixed point, and {$d_1$} is finite there. The reverse possibility---$\zeta>1{+\chi}$, {$d_1$} is irrelevant at the fixed point, and $f_0$ is finite there---is equally likely.
A third possibility, namely that {\it both} $f_0$  and {$d_1$} vanish at the fixed point, also exists.

Clearly, an exhaustive list of all these possibilities would be both exhaust{\it ing} and uninformative. The bottom line is that there is very little we can say about the scaling exponents $z$, $\zeta$, $\chi$, and $\chi_\rho$ for the immortal case, beyond the observation that they are {\it not} those predicted by the linear theory, and that they {\it are} universal. All claims to the contrary in the literature \cite{TT95,TT98,Solon_Chate, Ikeda,boost_agnostic}, including those by one of the current authors, are based on erroneous arguments, as we have argued above.

{Therefore, the only analytical assertions that we \emph{can} make with confidence are (i) the hydrodynamics of immortal flocks \emph{must} be modified by nonlinearities and (ii) $c_0$ \emph{cannot} be the \emph{only} relevant nonlinearity in $d=2$. If not for (ii), we could calculate the exponents exactly in $d=2$: $\chi=\chi_\rho=-1/5$, $\zeta=3/5$, $z=6/5$. However, as shown in Sec. \ref{rgmalt}, this fixed point is unstable to the $a_1$ nonlinearity. Unfortunately, this only rules out two sets of exponents out of infinite{ly many} possibilities.}

{\section{Summary and outlook}}
{In this work, we have comprehensively re-examined the hydrodynamic theory of flocks in two dimensions, }for both ``Malthusian'' (non-number-conserving) and ``immortal'' (number-conserving) systems. {For Malthusian flocks, w}e have 
obtained two exact scaling relations between the three scaling exponents that 
characterize the long-distance,  long-time behaviour of these systems.
We have argued that such flocks display long-range order in two dimensions.

We further demonstrate that for  \emph{immortal} flocks, there is no analytical argument that gives {\it any} exact scaling relations between the scaling exponents. This makes it impossible to demonstrate that such systems display long-range order in two dimensions. Therefore, the strongest evidence for the existence of long-range order in immortal flocks stems from extensive numerical studies \cite{chate2020dry}. 

{While the ``canonical'' exponents \cite{TT95, TT98} $z=2(d+1)/5$, $\zeta=(d+1)/5$, $\chi=(3-2d)/5$  are incorrect for flocks in $d=2$, they do hold for \emph{incompressible} flocks in $d\geq 3$ \cite{incmpdg2}.}

{Given that we have shown that we cannot obtain the exponents of flocks in $d=2$ analytically, is there any nonequilibrium flock whose hydrodynamic properties we \emph{can} calculate analytically in $d=2$? Yes: incompressible flocks in the presence of both annealed \cite{incomp_flockdg2, CLTM_anneal, AM_pol} and quenched disorder \cite{AM_disord, CLTM_disord1, CLTM_disord2}. While the long-range order in the former is due to the incompressibility constraint and not due to motion, the long-range order in the latter \emph{requires} motility. Similarly, we can exactly calculate the hydrodynamic properties of flocks at solid or fluid interfaces of bulk fluids \cite{Sarkar1, Sarkar2, AM_interface}.}
\\

\begin{acknowledgments} 
A.M. thanks Cesare Nardini for useful discussions and acknowledges the
support of ANR through the grant PSAM.  AM also acknowledges a TALENT fellowship awarded by CY Cergy Paris Universit\'e.  A.M. and
S.R. would like to thank the Isaac Newton Institute for
Mathematical Sciences, Cambridge, for support and hospitality
during the programmes ``New Statistical Physics in Living
Matter: non-equilibrium states under Adaptive Control'' and ``Anti-diffusive dynamics: from sub-cellular to astrophysical
scales'' (EPSRC grant no. EP/R014604/1), and SR additionally for a Rothschild Distinguished Visiting Fellowship. This research was supported in part by grant NSF PHY-2309135 to the Kavli Institute for Theoretical Physics. S.R. was supported in part by a J C Bose Fellowship of the ANRF, India. L.C. acknowledges support by the National Science Foundation
of China (under Grant No. 12274452), and thanks the MPI-PKS, where the early stage of this work was performed, for their support.
\end{acknowledgments}

	\bibliographystyle{apsrev4-2}
\bibliography{ref}

\end{document}